\documentclass[%
 reprint,
 superscriptaddress,
 amsmath,amssymb,
 physrev
]{revtex4-2}

\usepackage{graphicx}% Include figure files
\usepackage{dcolumn}% Align table columns on decimal point
\usepackage{bm}% bold math
\usepackage{savesym}
\savesymbol{tablenum}
\usepackage{siunitx}
\restoresymbol{SIX}{tablenum}

\usepackage{here}

\begin{document}

\preprint{}

\title{Interparticle normal force in highly porous granular matter during compression}% Force line breaks with \\
%%\thanks{A footnote to the article title}%

\author{Sota Arakawa}
\email{arakawas@jamstec.go.jp}
\affiliation{%
Yokohama Institute for Earth Sciences, Japan Agency for Marine-Earth Science and Technology, \\
3173-25, Showa-machi, Kanazawa-ku, Yokohama, 236-0001, Japan
}%

\author{Misako Tatsuuma}
\affiliation{
RIKEN Interdisciplinary Theoretical and Mathematical Sciences Program (iTHEMS), \\
2-1 Hirosawa, Wako, Saitama, 351-0198, Japan
}%
\affiliation{
Department of Earth and Planetary Sciences, Tokyo Institute of Technology, \\
2-12-1 Ookayama, Meguro, Tokyo, 152-8550, Japan
}%

\author{Hidekazu Tanaka}
\affiliation{
Astronomical Institute, Graduate School of Science, Tohoku University, \\
6-3 Aramaki, Aoba-ku, Sendai, 980-8578, Japan
}%

\author{Mikito Furuichi}%
\author{Daisuke Nishiura}%
\affiliation{%
Yokohama Institute for Earth Sciences, Japan Agency for Marine-Earth Science and Technology, \\
3173-25, Showa-machi, Kanazawa-ku, Yokohama, 236-0001, Japan
}%

%% \author{Naoya Sakatani}%
%% \author{Satoshi Tanaka}%
%% \affiliation{%
%% Institute of Space and Astronautical Science, Japan Aerospace Exploration Agency, \\
%% 3-1-1 Yoshinodai, Chuo-ku, Sagamihara, 252-5210, Japan
%% }%

\date{\today}% It is always \today, today,

\begin{abstract}

We perform a numerical simulation of compression of a highly porous dust aggregate of monodisperse spheres.
We find that the average interparticle normal force within the aggregate is inversely proportional to both the filling factor and the average coordination number, and we also derive this relation theoretically.
Our findings would be applicable for granular matter of arbitrary structures, as long as the
constituent particles are monodisperse spheres.

\end{abstract}

\maketitle

%\tableofcontents

\section{Introduction}

Granular materials are ubiquitous on earth and in space \cite{1996RvMP...68.1259J, 1997AREPS..25...85I, 2007Sci...316.1011M, 2008ARA&A..46...21B, 2011Sci...333.1125T}, and understanding their physical properties is of great importance in various fields of science and engineering \cite{2007NatPh...3..420K, Matsushima, CHEN2022117304}.
Granular materials have been usually defined as agglomerates of discrete particles, and the interparticle contact area is a function of the normal force acting between two contact particles \cite{hertz1896miscellaneous, 1971RSPSA.324..301J, 1975JCIS...53..314D}.
The interparticle contact area is a key parameter which controls thermal and mechanical properties of granular matter \cite{chan1973conductance, 1997ApJ...480..647D, GUSAROV20031103, 2017AIPA....7a5310S, 2017A&A...608L...7A}.
In addition, constituent particles would be broken when the interparticle force exceeds the threshold for failure \cite{2010PhRvL.104j8001B, 2015Icar..257...33S, 2022A&A...664A.147O}.
Therefore, the interparticle force in compressed granular matter has been intensively investigated \cite{1998PhRvE..57.3164M, 2001PhRvL..86..111O, 2005Natur.435.1079M}.

When constituent particles have large static frictions for tangential motions, highly porous structure would be achieved by compression with low pressure \cite{2009ApJ...701..130G, 2013A&A...554A...4K, 2017P&SS..149...14O}.
Indeed, highly porous dust aggregates with filling factors below 10\% might exist in protoplanetary disks as building blocks of planets \cite{2012ApJ...752..106O, 2013A&A...557L...4K, 2020Natur.579..518O, 2021ApJ...922...16K, 2023ApJ...944L..43T}.
In disks, such porous aggregates could be formed via pairwise collisional growth \cite{2000PhRvL..85.2426B, 2006Icar..182..274P, 2008ApJ...684.1310S}, and the initial structure of those aggregates would resemble that of fractal aggregates formed by ballistic cluster--cluster aggregation process \cite{WOS:000081472000002}.

However, the interparticle force in compressed fluffy aggregates has never been investigated.
This is because preparation of initial fractal aggregates in laboratories is difficult in nonzero gravity conditions \cite{2000Icar..143..138B}.
The measurement of interparticle force is also challenging when the force is small.
In contrast, numerical simulations are not affected by those difficulties, and we can investigate how the interparticle force changes with increasing the filling factor of porous aggregates.

In this study, we perform a three-dimensional numerical simulation using the soft-sphere discrete element method and demonstrate the temporal evolution of the interparticle force during omnidirectional compression.
As a consequence, we find that the average interparticle normal force, $\langle F \rangle$, is given by a simple function of the pressure, $P$, the filling factor, $\phi$, and the average coordination number, ${\langle Z \rangle}$.
We also reveal that this relation is directly derived from the definition of the pressure in granular matter.
Our findings would be widely applicable when we evaluate the thermal or mechanical properties of fluffy aggregates.

\section{Numerical Method}
\label{sec:method}

We perform a numerical simulation of compression of a highly porous dust aggregate of monodisperse spheres.
The number of particles in the simulation is $N = 2^{14} = 16384$, and the constituent dust particles are made of water ice whose radius is $r_{1} = 0.1~\si{\micro m}$.
The material properties including the elastic modulus and the surface energy are summarized in Ref.~\cite{2007ApJ...661..320W}.
The numerical code used in this study is identical to that of previous studies \cite{2019ApJ...874..159T, 2023ApJ...953....6T}.
We calculate the translational and rotational motions of each particle by solving the Newton--Euler equations.
We integrate these equations using the leapfrog method, which is a second-order symplectic integrator with a good accuracy of energy conservation.

In this study, we assume that the interparticle normal motion is described by a contact model for elastic cohesive spheres called the JKR model \cite{1971RSPSA.324..301J}.
The interparticle normal force, $F$, is a function of the compression length between two contact particles, $\delta$.
We define $\delta$ as
\begin{equation}
\delta = 2 r_{1} - d,
\end{equation}
and $d$ is the distance between the two particles’ centers.
Two contact particles make a circular contact area, and the contact radius, $a$, is also a function of $\delta$.

At the equilibrium state where $F = 0$, the contact radius is $a = a_{0}$ and the compression length is $\delta = \delta_{0}$.
Here $a_{0}$ is given by
\begin{equation}
a_{0} = {\left( \frac{9 \pi \gamma R^{2}}{E^{*}} \right)}^{1/3} = 12.4~\si{nm},
%% 1.2371201318108004e-8 
\end{equation}
where $\gamma = 0.1~\si{J.m^{-2}}$ \cite{2007ApJ...661..320W} is the surface energy, $E^{*}$ is the reduced Young’s modulus, and $R$ is the reduced particle radius \cite{1971RSPSA.324..301J}.
In this study, we assume that the two contact particles have the same radius and composition, and $R$ and $E^{*}$ are given by
\begin{equation}
R = \frac{r_{1}}{2},
\end{equation}
and
\begin{equation}
E^{*} = \frac{E}{2 {\left( 1 - \nu^{2} \right)}},
\end{equation}
where $E = 7~\si{GPa}$ is the Young’s modulus and $\nu = 0.25$ is the Poisson’s ratio \cite{2007ApJ...661..320W}.
The compression length at the equilibrium is
\begin{equation}
\delta_{0} = \frac{{a_{0}}^{2}}{3 R},
%% 1.020310813687715e-9
\end{equation}
and $\delta_{0} = 1.0~\si{nm}$ for $r_{1} = 0.1~\si{\micro m}$.

Here we introduce the normalized compression length, $x$, as follows:
\begin{equation}
x = \frac{\delta}{\delta_{0}}.
\end{equation}
We also introduce a dimensionless function, $y$, which describes the contact radius:
\begin{equation}
y = {\left( \frac{a}{a_{0}} \right)}^{1/2}.
\end{equation}
The relation between $x$ and $y$ is given by Ref.~\cite{1971RSPSA.324..301J} as follows:
\begin{equation}
3 y^{4} - 2 y - x = 0.
\label{eq:y-to-x}
\end{equation}
We also derive an equivalent equation which explicitly expresses $y$ as a function of $x$ as follows:
\begin{equation}
y = \frac{1}{2} {\left( - A {( x )} + \frac{4}{3 \sqrt{A {( x )}}} \right)}^{1/2} + \frac{\sqrt{A {( x )}}}{2},
\label{eq:x-to-y}
\end{equation}
where $A {( x )}$ is given by
\begin{equation}
A {( x )} = \frac{2^{1/3} \alpha {( x )}}{3} - \frac{2^{5/3} x}{3 \alpha {( x )}},
\end{equation}
and $\alpha {( x )}$ is
\begin{equation}
\alpha {( x )} = {\left( \sqrt{16 x^{3} + 9} + 3 \right)}^{1/3}.
\end{equation}
This is one of the real solution of Equation (\ref{eq:y-to-x}).
The interparticle contact breaks at $x = - {( 9 /16 )}^{1/3}$, and $y = {( 1 / 6 )}^{1/3}$ at the time.
For the JKR model, $y$ is a monotonically increasing function of $x$.
The explicit formulation derived here is useful when we calculate $y$ as a function of $x$ \footnote{
We note that an equivalent equation was derived in Ref.~\cite{CHEN2023118742} in a smart way.
}.

We also define the normalized force acting between two contact particles, $z$, as follows \cite{1971RSPSA.324..301J}:
\begin{equation}
z = \frac{F}{F_{\rm c}},
\end{equation}
where $F_{\rm c}$ is the maximum force needed to disconnect the two contact particles.
In the contact model of Ref.~\cite{1971RSPSA.324..301J}, $F_{\rm c}$ is given by
\begin{equation}
F_{\rm c} = 3 \pi \gamma R,
%% 4.712388980384689e-8
\end{equation}
and $F_{\rm c} = 4.7 \times 10^{-8}~\si{N}$ for $r_{1} = 0.1~\si{\micro m}$.
The relation between $y$ and $z$ is given by \cite{1971RSPSA.324..301J, 2007ApJ...661..320W}
\begin{equation}
z = 4 {\left( y^{6} - y^{3} \right)}.
\label{eq:y-to-z}
\end{equation}
By solving these equations, we can calculate $z$ as a function of $x$.
In other words, $F$ is given as a function of $\delta$.
We note that $F$ is positive when the repulsive force acts on two contact particles.

The elastic force in the normal direction induces oscillation.
However, in reality, the oscillation would be damped due to viscoelastic energy dissipation \cite{2012PThPS.195..101T, 2013JPhD...46Q5303K, 2021ApJ...910..130A}.
In our simulation, the damping force applied to each pair of two contact particles is introduced.
The detail of damping model is described in Ref.~\cite{2019ApJ...874..159T} (see Appendix \ref{app:damp}).

The interparticle tangential interactions are modeled by Ref.~\cite{2007ApJ...661..320W}.
We consider three types of motions: rolling, sliding, and twisting.
When the displacements are small, the resistances against these displacements are described by elastic spring models, while inelastic motions take place when the displacements exceed the critical values (see Figures 2 and 3 of Ref.~\cite{2007ApJ...661..320W}).
The detail of particle interaction models is described in Ref.~\cite{2007ApJ...661..320W} (see Appendix \ref{app:tangential}).

\begin{figure}[H]
\centering
\includegraphics[width=0.8\columnwidth]{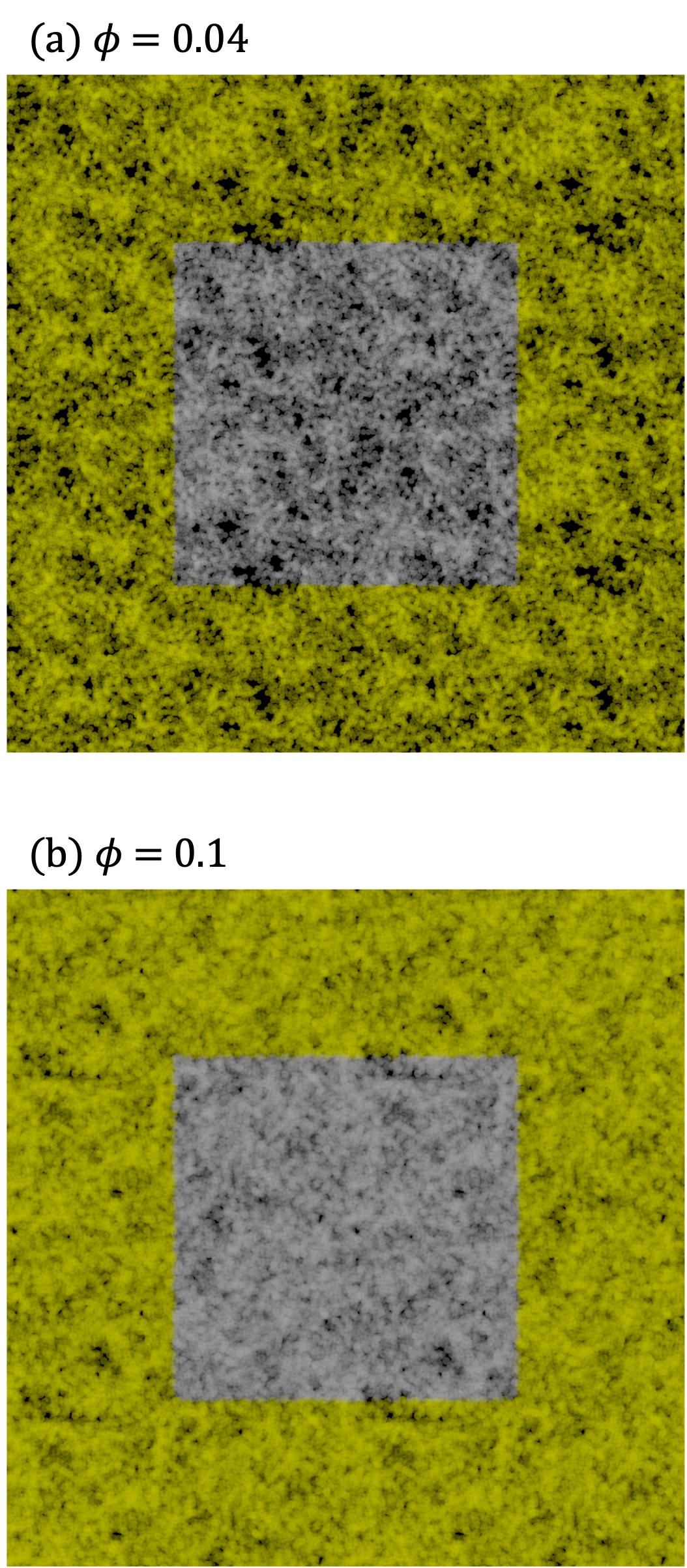}
\caption{\label{fig:1}
Snapshots of the structure evolution of an aggregate during compression in a cubic periodic boundary.
The gray particles are inside a cubic box.
We also plot the yellow particles which are in neighboring boxes to the box of gray particles.
The box size decreases with time and the filling factor, $\phi$, increases with time.
Panels (a) and (b) are the snapshots at $\phi = 0.04$ and $0.1$, respectively.
The length of the box is $L = 119.9 r_{1}$ for Panel (a) and $L = 88.2 r_{1}$ for Panel (b).
%% 119.905
%% 88.207
}
\end{figure}

We prepare an initial dust aggregate by ballistic cluster--cluster aggregation as in previous studies \cite{2013A&A...554A...4K, 2019ApJ...874..159T, 2023ApJ...953....6T}.
Then we perform an isotropic compression simulation as investigated by Refs.~\cite{2013A&A...554A...4K, 2023ApJ...953....6T}.
We adopt the periodic boundary condition (Figure \ref{fig:1}), and a cubic box with a volume of $L^{3}$ is considered as the computational region.
The box size decreases with time, $t$, as follows:
\begin{equation}
\frac{\mathrm{d} L}{\mathrm{d} t} = - \frac{2 C_{\rm v} L}{t_{\rm c}},
\label{eq:comp}
\end{equation}
where $C_{\rm v} = 1 \times 10^{-7}$ is the strain rate parameter \cite{2013A&A...554A...4K} and $t_{\rm c} = 1.93 \times 10^{-10}~\si{s}$ is the characteristic time of interparticle normal interaction \cite{2007ApJ...661..320W}.
The volume filling factor at each time step is defined as follows:
\begin{equation}
\phi = \frac{4 \pi {r_{1}}^{3} N}{3 L^{3}}.
\end{equation}

\section{Results and Discussion}
\label{sec:results}

\begin{figure}[H]
\centering
\includegraphics[width=\columnwidth]{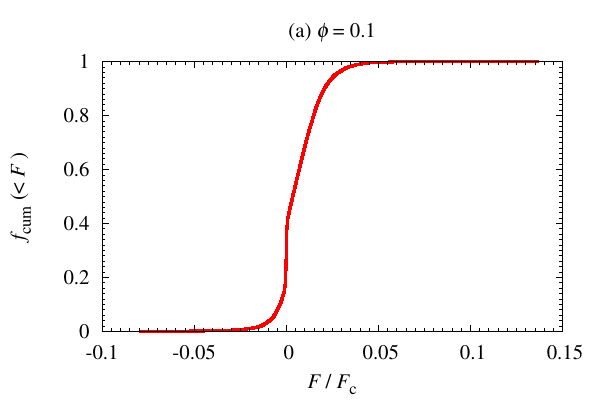}
\includegraphics[width=\columnwidth]{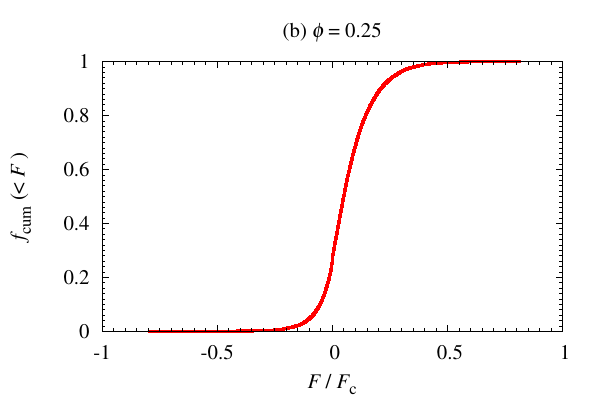}
\caption{\label{fig:freq}
Cumulative frequency distribution of interparticle normal force.
Here $f_{\rm cum} {(< F)}$ is the fraction of particle connections whose normal force is smaller than $F$.
The compression length between contacting particles is $\delta = \delta_{0}$ when $F = 0$.
Panels (a) and (b) are for the cases of $\phi = 0.1$ and $0.25$, respectively.
}
\end{figure}

First, we show the frequency distribution of interparticle normal force during compression \footnote{
Here we analyze all particle pairs with $x > 0$ as a post process analysis.
Strictly speaking, two particles in contact do not detach at $x = 0$ but the contact breaks at $x = - {(9 / 16)}^{1/3}$ in the JKR model.
However, in our compression simulation, the contribution of contacts with $x < 0$ should be negligibly small.
}.
Figure \ref{fig:freq} shows the cumulative frequency distribution of $F$ within an aggregate.
Here $f_{\rm cum} {(< F)}$ is the fraction of particle connections whose normal force is smaller than $F$.
The differential frequency distribution of $F$ is shown in Appendix \ref{app:diff} as a reference.

We can see a jump of $f_{\rm cum} {(< F)}$ at around $F = 0$ in Figure \ref{fig:freq}(a), and a large fraction of particle connections is piled up at $F \approx 0$ when $\phi \le 0.1$.
%% As shown in Figures \ref{fig:freq}(a) and \ref{fig:freq}(b), a large fraction of particle connections is piled up at $F \approx 0$ when $\phi \le 0.1$.
In contrast, no strong pile-up at $F \approx 0$ is observed for $\phi = 0.25$ (Figure \ref{fig:freq}(b)).
This result reflects the change of particle chain structures within the aggregate: a large fraction of particles in highly porous aggregates are not in backbone structure but in non-contributing dead-ends (see also Figure 6 of Ref.~\cite{2019PTEP.2019i3E02A}).
%% \cite{2017A&A...608L...7A}.

%% We note that the distribution of $f_{\rm cum} {(< F)}$ has long tails for both $F \ll 0$ and $F \gg 0$.
%% We interpret these tails reflect the oscillatory motion of constituent particles due to elastic interactions.
%% As the compression speed is nonzero in our simulation, a small fraction of particles oscillate during compression.
%% For the quasistatic limit, it would be natural that $f_{\rm cum} {(< F)}$ becomes zero for $F < 0$ when aggregates are compressed.
%% We expect that the fraction of these particles must depend on the compression speed and the strength of the damping force, and we will investigate these effects in future studies.
We investigate the dependence on the compression speed ($C_{\rm v}$) and damping force ($k_{\rm n}$) in Appendix \ref{app:slow}.
We confirm that the distribution of $f_{\rm cum} {(< F)}$ barely depends on $C_{\rm v}$ and $k_{\rm n}$.

\begin{figure}
\centering
\includegraphics[width=\columnwidth]{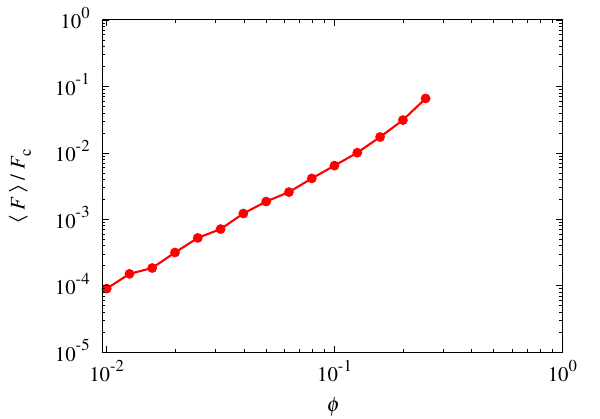}
\caption{\label{fig:F}
Average of the interparticle normal force, ${\langle F \rangle}$, as a function of $\phi$.
Here we assume that constituent particles are made of water ice and the radius is $r_{1} = 0.1~\si{\micro m}$.
}
\end{figure}

Next, we show the average of the interparticle normal force, ${\langle F \rangle}$, and its dependence on the filling factor. 
Figure \ref{fig:F} shows ${\langle F \rangle}$ as a function of $\phi$.
%% We find that ${\langle F \rangle}$ increases with $\phi$ during compression, and ${\langle F \rangle} / F_{\rm c} \ll 1$ for $\phi \le 0.25$.
%% Then we can roughly regard $\delta$ for each particle connection as constant and $\delta \approx \delta_{0}$.
We find that ${\langle F \rangle}$ increases with $\phi$ during compression, and for $\phi \le 0.25$ we find ${\langle F \rangle} / F_{\rm c} \ll 1$ which allows us to consider the compression length to be constant, $\delta \approx \delta_{0}$.
We note that ${\langle F \rangle} / F_{\rm c}$ also depends on the material properties and radius of constituent particles, and this result is only applicable for aggregates made of water ice particles with $r_{1} = 0.1~\si{\micro m}$.

Finally, we discuss the relation between ${\langle F \rangle}$ and the pressure of the dust aggregate, $P$.
In this study, an aggregate is compressed by themselves as we use the periodic boundary condition.
The pressure of the aggregate is defined by the standard method in molecular dynamics simulations which is based on the virial theorem \cite{haile1997molecular}.
In our numerical simulation, $P$ is defined as follows \cite{2013A&A...554A...4K, o2011particulate}:
\begin{equation}
P = {\Bigg\langle \frac{2}{3 L^{3}} \sum_{i = 1}^{N} \frac{m {v_{i}}^{2}}{2} + \frac{1}{3 L^{3}} \sum_{i < j} d_{i, j} F_{i, j} \Bigg\rangle}_{t},
\label{eq:P_def}
\end{equation}
where $m$ is the mass of each particle, $v_{i}$ is the velocity of the $i$th particle, $d_{i, j}$ is the distance between the $i$th and $j$th particles, and $F_{i, j}$ is the the interparticle normal force between $i$th and $j$th particles.
Assuming that the material density of ice is $\rho = 1000~\si{kg.m^{-3}}$ \cite{2007ApJ...661..320W}, $m$ is given by $m = {( 4 \pi / 3 )} \rho {r_{1}}^{3} = 4.2 \times 10^{-18}~\si{kg}$.
Here $F_{i, j}$ is positive when the repulsive force works, and $F_{i, j} = 0$ if $i$th and $j$th particles do not contact.
We calculate the time-averaged value of $P$, and ${\langle \mathcal{A} \rangle}_{t}$ denotes the time average of a variable $\mathcal{A}$.
We take an average of the right-hand side of Equation (\ref{eq:P_def}) for $10^{3} t_{\rm c}$, which is sufficiently longer than the characteristic time of particle interaction ($= t_{\rm c}$) and negligibly shorter than the timescale of compression ($= 5 \times 10^{6} t_{\rm c}$; see Equation (\ref{eq:comp})) \cite{2013A&A...554A...4K, 2023ApJ...953....6T}.

\begin{figure}[H]
\centering
\includegraphics[width=\columnwidth]{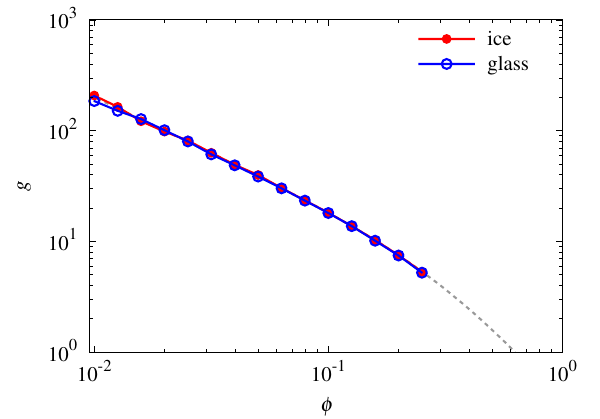}
\caption{\label{fig:FP}
Normalized normal force, $g$, as a function of $\phi$.
The dashed line is a theoretical prediction (Equation (\ref{eq:g_pre})).
Red and blue points denote our numerical result for ice and glass spheres, respectively.
}
\end{figure}

We introduce the normalized normal force, $g$, as follows: 
\begin{equation}
g = \frac{\langle F \rangle}{\pi {r_{1}}^{2} P}.
\label{eq:g_def}
\end{equation}
In our simulation, the porous aggregate is continuously compressed and the applied pressure is balanced with the compressive strength.
In contrast, when the applied pressure is lower than the compressive strength and the deformation of an aggregate is negligible, ${\langle F \rangle}$ must be proportional to $P$.
Thus, we can interpret $g$ as a constant of proportionality.
The red line of Figure \ref{fig:FP} shows $g$ for the range between $\phi = 0.01$ and $0.25$.

It should be noted that $g$ would be independent of the material parameters of the constituent particles.
Here we perform an additional simulation using dust aggregates of glass spheres.
The material parameters of glass spheres are $\gamma = 0.02~\si{J.m^{-2}}$, $E = 54~\si{GPa}$, $\nu = 0.17$, and $\rho = 2650~\si{kg.m^{-3}}$ \cite{2007ApJ...661..320W}.
We set $C_{\rm v} = 3 \times 10^{-7}$ and $k_{\rm n} = 0.1$ for this case.
The particle radius is set to be equal to that for ice aggregates: $r_{1} = 0.1~\si{\micro m}$.
The number of constituent particle and the initial structure of the aggregate are also identical to those for ice aggregates.
The blue line of Figure \ref{fig:FP} denote our numerical result for glass spheres.
It is obvious that numerical results for both ice and glass aggregates are consistent with each other.

Here we show that $g$ could be derived analytically from the definition of $P$.
When the compression speed is sufficiently low, we can regard the compression as a quasistatic process (i.e., $v_{i} \approx 0$), and Equation (\ref{eq:P_def}) is approximated by
\begin{equation}
P \approx \frac{1}{3 L^{3}} \sum_{i < j} d_{i, j} F_{i, j}.
\end{equation}
When ${\langle F \rangle} / F_{\rm c} \ll 1$, we can assume that $d_{i, j} \approx 2 r_{1} - \delta_{0}$ for all particle connections, and we obtain the following equation:
\begin{equation}
\sum_{i < j} d_{i, j} F_{i, j} \approx \frac{2 r_{1}}{c_{0}} {\langle F \rangle} N \frac{\langle Z \rangle}{2},
\end{equation}
where
\begin{equation}
c_{0} = \frac{2 r_{1}}{2 r_{1} - \delta_{0}}
\end{equation}
is a correction factor ($c_{0} = 1.005$ for ice particles with $r_{1} = 0.1~\si{\micro m}$), and ${\langle Z \rangle}$ is the average coordination number.
Therefore, we derive the following equation:
\begin{equation}
g \approx \frac{4 c_{0}}{{\langle Z \rangle} \phi},
\label{eq:g_pre}
\end{equation}
and $g$ is inversely proportional to both $\phi$ and ${\langle Z \rangle}$.
The dashed line in Figure \ref{fig:FP} is the theoretical prediction, and it shows excellent agreement with the numerical result.

\begin{figure}[H]
\centering
\includegraphics[width=\columnwidth]{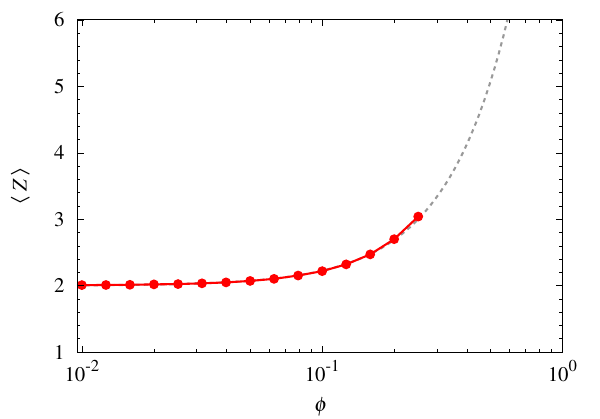}
\caption{\label{fig:Z}
Average coordination number, ${\langle Z \rangle}$, as a function of $\phi$.
The dashed line is a semi-analytic model in previous studies: ${\langle Z \rangle} = 2 + 9.38 \phi^{1.62}$ \cite{2019Icar..324....8A, 2019PTEP.2019i3E02A}.
}
\end{figure}

The average coordination number should also be a function of $\phi$, and it must depend on how to prepare the initial aggregate before compression.
Figure \ref{fig:Z} shows the filling factor dependence of ${\langle Z \rangle}$.
When the initial fluffy aggregates were prepared by ballistic cluster--cluster aggregation process, the filling factor dependence of ${\langle Z \rangle}$ is given by \cite{2019Icar..324....8A, 2019PTEP.2019i3E02A}
\begin{equation}
{\langle Z \rangle} = 2 + 9.38 \phi^{1.62}.
\end{equation}
We confirm that our numerical result is consistent with a model prediction.

Although the filling factor dependence of $g$ is trivial when we go back to the definition of $P$ in molecular dynamics simulations, it has been poorly understood.
Several predictions on $g$ have been reported; e.g.,  $g = {( 2 / \sqrt{6} )} \phi^{-1}$ \cite{2017AIPA....7a5310S} or $g = \phi^{-1}$ \cite{2015Icar..257...33S}.
These studies did not mention the dependence of $g$ on ${\langle Z \rangle}$, however.
When we focus on fluffy aggregates with $\phi \ll 0.1$, we can regard ${\langle Z \rangle} \approx 2$ and it barely depends on $\phi$.
In contrast, for $\phi \gtrsim 0.1$, ${\langle Z \rangle}$ clearly depends on $\phi$ (Figure \ref{fig:Z}) and the effect of ${\langle Z \rangle}$ on $g$ is non-negligible.

We note that Equation (\ref{eq:g_pre}) would be applicable for dust aggregates of arbitrary structures, as long as the constituent particles could be regarded as monodisperse spheres.
We expect that ${\langle Z \rangle} \approx 2$ whenever $\phi$ is order(s) of magnitude lower than $1$, however, the filling factor dependence of ${\langle Z \rangle}$ is different for each preparation procedure of aggregates \cite{2019PTEP.2019i3E02A, 2013A&A...551A..65S}.

It should be noted that not all particles do not contribute heat and pressure transfer within aggregates.
As shown in Figure \ref{fig:freq}, a substantial fraction of interparticle contacts is force free when $\phi \le 0.1$.
We might need to evaluate the effective mean of $F$ by averaging over contacts with nonzero $F$ when the effects on the thermal and mechanical properties of aggregates are considered.
%% To discuss the fraction of non-contributing contacts, we need to suppress the oscillatory motion of constituent particles.
%% We will perform numerical simulations with sufficiently low compression speed in future studies.

\section{Conclusion}

Summarizing, the relationship among the pressure within an aggregate, $P$, the filling factor, $\phi$, the average coordination number, ${\langle Z \rangle}$, and the average interparticle normal force, ${\langle F \rangle}$, is derived once numerically and then theoretically (Equations (\ref{eq:g_def}) and (\ref{eq:g_pre})).
We found that ${\langle F \rangle}$ is inversely proportional to both $\phi$ and ${\langle Z \rangle}$.
The filling factor dependence is consistent with that predicted in previous studies \cite{2017AIPA....7a5310S, 2015Icar..257...33S}.
Our findings of the dependence on ${\langle Z \rangle}$ would be new, and we can derive this dependence from the definition of the pressure in granular matter. 
We also note that $g$ would be independent of the material parameters of the constituent particles (Figure \ref{fig:FP}).

Understanding the interparticle normal force and its dependence on the other parameters are essential to predict the thermal and mechanical properties of granular matter.
Our findings will provide deeper insight into the physics of porous granular matter.
We expect that our theoretical prediction will be tested by laboratory experiments.

Finally, we note that not only the average of $F$ but also the distribution of $F$ is of great interest.
The failure of particles under pressure should be start when the maximum of $F$ exceeds the threshold, and the disruption of constituent particles changes the size distribution of particles.
The force chain structure should also be affected by the failure.
We will address these issues in future studies.

\section*{Acknowledgments}

This work was supported by JSPS KAKENHI grant Nos.~JP22J00260 and JP22KJ1292.

\appendix

\section{Damping force for interparticle normal motion}
\label{app:damp}

The elastic interparticle normal force, $F$, induces oscillation at each connection.
The oscillation would attenuate in reality due to energy dissipation.
In this study, we introduce an artificial damping force in the normal direction which is modeled in Ref.~\cite{2013A&A...554A...4K}.
The damping force applied to each particle is given by
\begin{equation}
F_{\rm damp} = - k_{\rm n} \frac{m}{t_{\rm c}} v_{\rm rel, n},
\end{equation}
where $k_{\rm n}$ is the dimensionless coefficient and $v_{\rm rel, n}$ is the normal component of the relative velocity between two contacting particles.
Note that $v_{\rm rel, n}$ is negative when two particles approach.
We adopt $k_{\rm n} = 0.01$ as a fiducial value \cite{2023ApJ...953....6T}.
%% We also note that the filling factor dependence of $P$ is nearly independent of the choice of $k_{\rm n}$ when the strain rate parameter is sufficiently small ($C_{\rm v} \le 3 \times 10^{-7}$) \cite{2013A&A...554A...4K}.

\section{Tangential interaction models}
\label{app:tangential}

We calculate the interaction of each connection of particles, taking all interactions modeled by Ref.~\cite{2007ApJ...661..320W} into account.
The mechanical model for the normal interaction is described in Section \ref{sec:method} in detail.
Here we briefly explain the models for the tangential interactions.

We consider three types of tangential motions, namely, rolling, sliding, and twisting (see Figure 2 of Ref.~\cite{2007ApJ...661..320W}).
The displacements corresponding to these motions are expressed by the rotation of the two particles in contact.
In the framework of the contact model developed by Ref.~\cite{2007ApJ...661..320W}, the elastic and inelastic regimes are considered for each interaction: no energy is dissipated when the displacements of the tangential motions are all small enough, while energy dissipation occurs when the displacements exceed the threshold values.
The forces and torques on each particle due to tangential interactions are originally formulated by Refs.~\cite{1995PMagA..72..783D, 1996PMagA..73.1279D}.
The detail of the interaction models is described in Section 2.2 of Ref.~\cite{2007ApJ...661..320W}.

\section{Differential frequency distribution of interparticle normal force}
\label{app:diff}

In the granular community, the frequency distribution of interparticle normal force is usually presented in the differential frequency distribution \cite{1998PhRvE..57.3164M, 2001PhRvL..86..111O, 2005Natur.435.1079M}.
Although we choose the cumulative distribution instead of the differential one in Section \ref{sec:results}, we present the differential frequency distribution in this appendix.

\begin{figure}[H]
\centering
\includegraphics[width=\columnwidth]{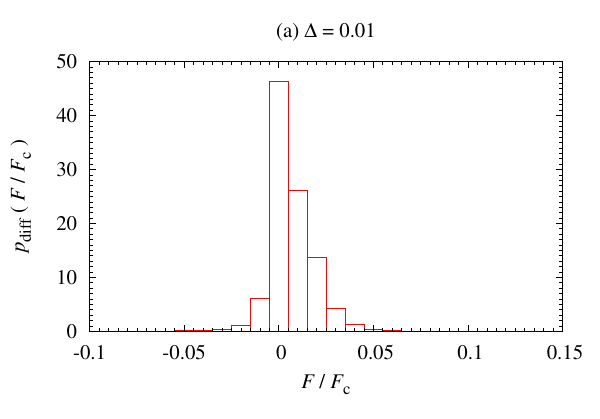}
\includegraphics[width=\columnwidth]{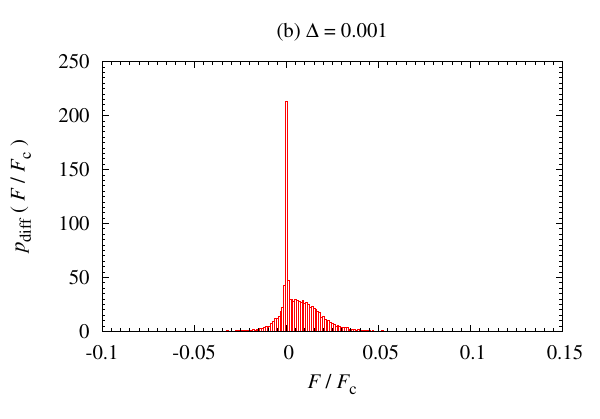}
\caption{\label{fig:freq_p}
Differential frequency distribution of interparticle normal force for the case of $\phi = 0.1$ (see Figure \ref{fig:freq}(b)).
Panels (a) and (b) are for the cases of $\Delta = 0.01$ and $0.001$, respectively.
}
\end{figure}

Figure \ref{fig:freq_p} shows the differential frequency distribution of interparticle normal force, $p_{\rm diff} {( F / F_{\rm c} )}$, for the case of $\phi = 0.1$ (see Figure \ref{fig:freq}(b)).
Here $p_{\rm diff} {( F / F_{\rm c} )}$ is given by
\begin{eqnarray}
p_{\rm diff} {( F / F_{\rm c} )} & = & \frac{f_{\rm cum} {(< {( i + 1/2 )} \Delta F_{\rm c} )}}{\Delta} \nonumber \\
& & - \frac{f_{\rm cum} {(< {( i - 1/2 )} \Delta F_{\rm c} )}}{\Delta},
\end{eqnarray}
where $\Delta$ is the bin width of the differential distribution, and $F = i \Delta F_{\rm c}$ ($i = 0, \pm 1, \pm 2$, ...).
Panels (a) and (b) are for the cases of $\Delta = 0.01$ and $0.001$, respectively. 
We find that the shape of the distribution is strongly affected by the choice of $\Delta$.

We can see a clear pile-up at $F \approx 0$ in Figure \ref{fig:freq_p}(b), as discussed in Section \ref{sec:results} (see also Figure \ref{fig:freq}(b)).
This reflects the fact that a non-negligible fraction of particles in highly porous aggregates are not in backbone structure.

\section{Dependence on $C_{\rm v}$ and $k_{\rm n}$}
\label{app:slow}

\begin{figure}[H]
\centering
\includegraphics[width=\columnwidth]{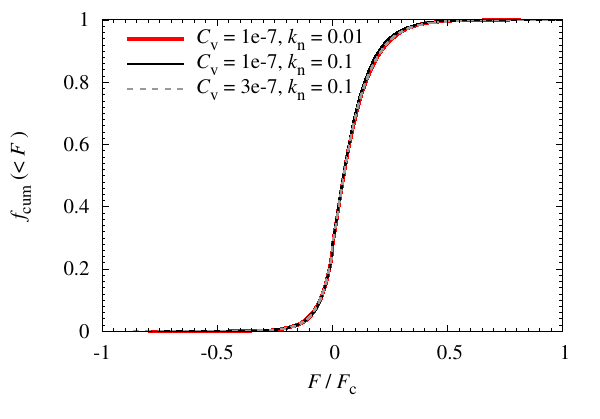}
\caption{\label{fig:freq_fast}
Cumulative frequency distribution of interparticle normal force, $f_{\rm cum} {(< F)}$, for $\phi = 0.25$.
The red line represents the fiducial case ($C_{\rm v} = 1 \times 10^{-7}$ and $k_{\rm n} = 0.01$).
The black line is a strong dissipation case ($C_{\rm v} = 1 \times 10^{-7}$ and $k_{\rm n} = 0.1$).
The gray dashed line is a strong dissipation and fast compression case ($C_{\rm v} = 3 \times 10^{-7}$ and $k_{\rm n} = 0.1$).
}
\end{figure}

To check the robustness of our numerical results, we perform additional simulations with different parameter sets of $C_{\rm v}$ and $k_{\rm n}$.
Figure \ref{fig:freq_fast} shows the cumulative frequency distribution of interparticle normal force, $f_{\rm cum} {(< F)}$, for $\phi = 0.25$.
The red line represents the fiducial case, and black and gray dashed lines show the results for different sets of $C_{\rm v}$ and $k_{\rm n}$.
We confirm that $f_{\rm cum} {(< F)}$ is approximately independent of $C_{\rm v}$ and $k_{\rm n}$.

\nocite{*}

\bibliography{apssamp}% Produces the bibliography via BibTeX.

\providecommand{\noopsort}[1]{}\providecommand{\singleletter}[1]{#1}%
\begin{thebibliography}{51}%
\makeatletter
\providecommand \@ifxundefined [1]{%
 \@ifx{#1\undefined}
}%
\providecommand \@ifnum [1]{%
 \ifnum #1\expandafter \@firstoftwo
 \else \expandafter \@secondoftwo
 \fi
}%
\providecommand \@ifx [1]{%
 \ifx #1\expandafter \@firstoftwo
 \else \expandafter \@secondoftwo
 \fi
}%
\providecommand \natexlab [1]{#1}%
\providecommand \enquote  [1]{``#1''}%
\providecommand \bibnamefont  [1]{#1}%
\providecommand \bibfnamefont [1]{#1}%
\providecommand \citenamefont [1]{#1}%
\providecommand \href@noop [0]{\@secondoftwo}%
\providecommand \href [0]{\begingroup \@sanitize@url \@href}%
\providecommand \@href[1]{\@@startlink{#1}\@@href}%
\providecommand \@@href[1]{\endgroup#1\@@endlink}%
\providecommand \@sanitize@url [0]{\catcode `\\12\catcode `\$12\catcode
  `\&12\catcode `\#12\catcode `\^12\catcode `\_12\catcode `\%12\relax}%
\providecommand \@@startlink[1]{}%
\providecommand \@@endlink[0]{}%
\providecommand \url  [0]{\begingroup\@sanitize@url \@url }%
\providecommand \@url [1]{\endgroup\@href {#1}{\urlprefix }}%
\providecommand \urlprefix  [0]{URL }%
\providecommand \Eprint [0]{\href }%
\providecommand \doibase [0]{https://doi.org/}%
\providecommand \selectlanguage [0]{\@gobble}%
\providecommand \bibinfo  [0]{\@secondoftwo}%
\providecommand \bibfield  [0]{\@secondoftwo}%
\providecommand \translation [1]{[#1]}%
\providecommand \BibitemOpen [0]{}%
\providecommand \bibitemStop [0]{}%
\providecommand \bibitemNoStop [0]{.\EOS\space}%
\providecommand \EOS [0]{\spacefactor3000\relax}%
\providecommand \BibitemShut  [1]{\csname bibitem#1\endcsname}%
\let\auto@bib@innerbib\@empty
%</preamble>
\bibitem [{\citenamefont {{Jaeger}}\ \emph {et~al.}(1996)\citenamefont
  {{Jaeger}}, \citenamefont {{Nagel}},\ and\ \citenamefont
  {{Behringer}}}]{1996RvMP...68.1259J}%
  \BibitemOpen
  \bibfield  {author} {\bibinfo {author} {\bibfnamefont {H.~M.}\ \bibnamefont
  {{Jaeger}}}, \bibinfo {author} {\bibfnamefont {S.~R.}\ \bibnamefont
  {{Nagel}}},\ and\ \bibinfo {author} {\bibfnamefont {R.~P.}\ \bibnamefont
  {{Behringer}}},\ }\bibfield  {title} {\bibinfo {title} {{Granular solids,
  liquids, and gases}},\ }\href {https://doi.org/10.1103/RevModPhys.68.1259}
  {\bibfield  {journal} {\bibinfo  {journal} {Reviews of Modern Physics}\
  }\textbf {\bibinfo {volume} {68}},\ \bibinfo {pages} {1259} (\bibinfo {year}
  {1996})}\BibitemShut {NoStop}%
\bibitem [{\citenamefont {{Iverson}}\ \emph {et~al.}(1997)\citenamefont
  {{Iverson}}, \citenamefont {{Reid}},\ and\ \citenamefont
  {{Lahusen}}}]{1997AREPS..25...85I}%
  \BibitemOpen
  \bibfield  {author} {\bibinfo {author} {\bibfnamefont {R.~M.}\ \bibnamefont
  {{Iverson}}}, \bibinfo {author} {\bibfnamefont {M.~E.}\ \bibnamefont
  {{Reid}}},\ and\ \bibinfo {author} {\bibfnamefont {R.~G.}\ \bibnamefont
  {{Lahusen}}},\ }\bibfield  {title} {\bibinfo {title} {{Debris-Flow
  Mobilization from Landslides}},\ }\href
  {https://doi.org/10.1146/annurev.earth.25.1.85} {\bibfield  {journal}
  {\bibinfo  {journal} {Annual Review of Earth and Planetary Sciences}\
  }\textbf {\bibinfo {volume} {25}},\ \bibinfo {pages} {85} (\bibinfo {year}
  {1997})}\BibitemShut {NoStop}%
\bibitem [{\citenamefont {{Miyamoto}}\ \emph {et~al.}(2007)\citenamefont
  {{Miyamoto}}, \citenamefont {{Yano}}, \citenamefont {{Scheeres}},
  \citenamefont {{Abe}}, \citenamefont {{Barnouin-Jha}}, \citenamefont
  {{Cheng}}, \citenamefont {{Demura}}, \citenamefont {{Gaskell}}, \citenamefont
  {{Hirata}}, \citenamefont {{Ishiguro}}, \citenamefont {{Michikami}},
  \citenamefont {{Nakamura}}, \citenamefont {{Nakamura}}, \citenamefont
  {{Saito}},\ and\ \citenamefont {{Sasaki}}}]{2007Sci...316.1011M}%
  \BibitemOpen
  \bibfield  {author} {\bibinfo {author} {\bibfnamefont {H.}~\bibnamefont
  {{Miyamoto}}}, \bibinfo {author} {\bibfnamefont {H.}~\bibnamefont {{Yano}}},
  \bibinfo {author} {\bibfnamefont {D.~J.}\ \bibnamefont {{Scheeres}}},
  \bibinfo {author} {\bibfnamefont {S.}~\bibnamefont {{Abe}}}, \bibinfo
  {author} {\bibfnamefont {O.}~\bibnamefont {{Barnouin-Jha}}}, \bibinfo
  {author} {\bibfnamefont {A.~F.}\ \bibnamefont {{Cheng}}}, \bibinfo {author}
  {\bibfnamefont {H.}~\bibnamefont {{Demura}}}, \bibinfo {author}
  {\bibfnamefont {R.~W.}\ \bibnamefont {{Gaskell}}}, \bibinfo {author}
  {\bibfnamefont {N.}~\bibnamefont {{Hirata}}}, \bibinfo {author}
  {\bibfnamefont {M.}~\bibnamefont {{Ishiguro}}}, \bibinfo {author}
  {\bibfnamefont {T.}~\bibnamefont {{Michikami}}}, \bibinfo {author}
  {\bibfnamefont {A.~M.}\ \bibnamefont {{Nakamura}}}, \bibinfo {author}
  {\bibfnamefont {R.}~\bibnamefont {{Nakamura}}}, \bibinfo {author}
  {\bibfnamefont {J.}~\bibnamefont {{Saito}}},\ and\ \bibinfo {author}
  {\bibfnamefont {S.}~\bibnamefont {{Sasaki}}},\ }\bibfield  {title} {\bibinfo
  {title} {{Regolith Migration and Sorting on Asteroid Itokawa}},\ }\href
  {https://doi.org/10.1126/science.1134390} {\bibfield  {journal} {\bibinfo
  {journal} {Science}\ }\textbf {\bibinfo {volume} {316}},\ \bibinfo {pages}
  {1011} (\bibinfo {year} {2007})}\BibitemShut {NoStop}%
\bibitem [{\citenamefont {{Blum}}\ and\ \citenamefont
  {{Wurm}}(2008)}]{2008ARA&A..46...21B}%
  \BibitemOpen
  \bibfield  {author} {\bibinfo {author} {\bibfnamefont {J.}~\bibnamefont
  {{Blum}}}\ and\ \bibinfo {author} {\bibfnamefont {G.}~\bibnamefont
  {{Wurm}}},\ }\bibfield  {title} {\bibinfo {title} {{The growth mechanisms of
  macroscopic bodies in protoplanetary disks.}},\ }\href
  {https://doi.org/10.1146/annurev.astro.46.060407.145152} {\bibfield
  {journal} {\bibinfo  {journal} {Annual Review of Astronomy and Astrophysics}\
  }\textbf {\bibinfo {volume} {46}},\ \bibinfo {pages} {21} (\bibinfo {year}
  {2008})}\BibitemShut {NoStop}%
\bibitem [{\citenamefont {{Tsuchiyama}}\ \emph {et~al.}(2011)\citenamefont
  {{Tsuchiyama}}, \citenamefont {{Uesugi}}, \citenamefont {{Matsushima}},
  \citenamefont {{Michikami}}, \citenamefont {{Kadono}}, \citenamefont
  {{Nakamura}}, \citenamefont {{Uesugi}}, \citenamefont {{Nakano}},
  \citenamefont {{Sandford}}, \citenamefont {{Noguchi}}, \citenamefont
  {{Matsumoto}}, \citenamefont {{Matsuno}}, \citenamefont {{Nagano}},
  \citenamefont {{Imai}}, \citenamefont {{Takeuchi}}, \citenamefont {{Suzuki}},
  \citenamefont {{Ogami}}, \citenamefont {{Katagiri}}, \citenamefont
  {{Ebihara}}, \citenamefont {{Ireland}}, \citenamefont {{Kitajima}},
  \citenamefont {{Nagao}}, \citenamefont {{Naraoka}}, \citenamefont
  {{Noguchi}}, \citenamefont {{Okazaki}}, \citenamefont {{Yurimoto}},
  \citenamefont {{Zolensky}}, \citenamefont {{Mukai}}, \citenamefont {{Abe}},
  \citenamefont {{Yada}}, \citenamefont {{Fujimura}}, \citenamefont
  {{Yoshikawa}},\ and\ \citenamefont {{Kawaguchi}}}]{2011Sci...333.1125T}%
  \BibitemOpen
  \bibfield  {author} {\bibinfo {author} {\bibfnamefont {A.}~\bibnamefont
  {{Tsuchiyama}}}, \bibinfo {author} {\bibfnamefont {M.}~\bibnamefont
  {{Uesugi}}}, \bibinfo {author} {\bibfnamefont {T.}~\bibnamefont
  {{Matsushima}}}, \bibinfo {author} {\bibfnamefont {T.}~\bibnamefont
  {{Michikami}}}, \bibinfo {author} {\bibfnamefont {T.}~\bibnamefont
  {{Kadono}}}, \bibinfo {author} {\bibfnamefont {T.}~\bibnamefont
  {{Nakamura}}}, \bibinfo {author} {\bibfnamefont {K.}~\bibnamefont
  {{Uesugi}}}, \bibinfo {author} {\bibfnamefont {T.}~\bibnamefont {{Nakano}}},
  \bibinfo {author} {\bibfnamefont {S.~A.}\ \bibnamefont {{Sandford}}},
  \bibinfo {author} {\bibfnamefont {R.}~\bibnamefont {{Noguchi}}}, \bibinfo
  {author} {\bibfnamefont {T.}~\bibnamefont {{Matsumoto}}}, \bibinfo {author}
  {\bibfnamefont {J.}~\bibnamefont {{Matsuno}}}, \bibinfo {author}
  {\bibfnamefont {T.}~\bibnamefont {{Nagano}}}, \bibinfo {author}
  {\bibfnamefont {Y.}~\bibnamefont {{Imai}}}, \bibinfo {author} {\bibfnamefont
  {A.}~\bibnamefont {{Takeuchi}}}, \bibinfo {author} {\bibfnamefont
  {Y.}~\bibnamefont {{Suzuki}}}, \bibinfo {author} {\bibfnamefont
  {T.}~\bibnamefont {{Ogami}}}, \bibinfo {author} {\bibfnamefont
  {J.}~\bibnamefont {{Katagiri}}}, \bibinfo {author} {\bibfnamefont
  {M.}~\bibnamefont {{Ebihara}}}, \bibinfo {author} {\bibfnamefont {T.~R.}\
  \bibnamefont {{Ireland}}}, \bibinfo {author} {\bibfnamefont {F.}~\bibnamefont
  {{Kitajima}}}, \bibinfo {author} {\bibfnamefont {K.}~\bibnamefont {{Nagao}}},
  \bibinfo {author} {\bibfnamefont {H.}~\bibnamefont {{Naraoka}}}, \bibinfo
  {author} {\bibfnamefont {T.}~\bibnamefont {{Noguchi}}}, \bibinfo {author}
  {\bibfnamefont {R.}~\bibnamefont {{Okazaki}}}, \bibinfo {author}
  {\bibfnamefont {H.}~\bibnamefont {{Yurimoto}}}, \bibinfo {author}
  {\bibfnamefont {M.~E.}\ \bibnamefont {{Zolensky}}}, \bibinfo {author}
  {\bibfnamefont {T.}~\bibnamefont {{Mukai}}}, \bibinfo {author} {\bibfnamefont
  {M.}~\bibnamefont {{Abe}}}, \bibinfo {author} {\bibfnamefont
  {T.}~\bibnamefont {{Yada}}}, \bibinfo {author} {\bibfnamefont
  {A.}~\bibnamefont {{Fujimura}}}, \bibinfo {author} {\bibfnamefont
  {M.}~\bibnamefont {{Yoshikawa}}},\ and\ \bibinfo {author} {\bibfnamefont
  {J.}~\bibnamefont {{Kawaguchi}}},\ }\bibfield  {title} {\bibinfo {title}
  {{Three-Dimensional Structure of Hayabusa Samples: Origin and Evolution of
  Itokawa Regolith}},\ }\href {https://doi.org/10.1126/science.1207807}
  {\bibfield  {journal} {\bibinfo  {journal} {Science}\ }\textbf {\bibinfo
  {volume} {333}},\ \bibinfo {pages} {1125} (\bibinfo {year}
  {2011})}\BibitemShut {NoStop}%
\bibitem [{\citenamefont {{Katsuragi}}\ and\ \citenamefont
  {{Durian}}(2007)}]{2007NatPh...3..420K}%
  \BibitemOpen
  \bibfield  {author} {\bibinfo {author} {\bibfnamefont {H.}~\bibnamefont
  {{Katsuragi}}}\ and\ \bibinfo {author} {\bibfnamefont {D.~J.}\ \bibnamefont
  {{Durian}}},\ }\bibfield  {title} {\bibinfo {title} {{Unified force law for
  granular impact cratering}},\ }\href {https://doi.org/10.1038/nphys583}
  {\bibfield  {journal} {\bibinfo  {journal} {Nature Physics}\ }\textbf
  {\bibinfo {volume} {3}},\ \bibinfo {pages} {420} (\bibinfo {year}
  {2007})}\BibitemShut {NoStop}%
\bibitem [{\citenamefont {{Matsushima}}\ \emph {et~al.}(2009)\citenamefont
  {{Matsushima}}, \citenamefont {{Katagiri}}, \citenamefont {{Uesugi}},
  \citenamefont {{Tsuchiyama}},\ and\ \citenamefont {{Nakano}}}]{Matsushima}%
  \BibitemOpen
  \bibfield  {author} {\bibinfo {author} {\bibfnamefont {T.}~\bibnamefont
  {{Matsushima}}}, \bibinfo {author} {\bibfnamefont {J.}~\bibnamefont
  {{Katagiri}}}, \bibinfo {author} {\bibfnamefont {K.}~\bibnamefont
  {{Uesugi}}}, \bibinfo {author} {\bibfnamefont {A.}~\bibnamefont
  {{Tsuchiyama}}},\ and\ \bibinfo {author} {\bibfnamefont {T.}~\bibnamefont
  {{Nakano}}},\ }\bibfield  {title} {\bibinfo {title} {{3D Shape
  Characterization and Image-Based DEM Simulation of the Lunar Soil Simulant
  FJS-1}},\ }\href {https://doi.org/10.1061/(ASCE)0893-1321(2009)22:1(15)}
  {\bibfield  {journal} {\bibinfo  {journal} {Journal of Aerospace
  Engineering}\ }\textbf {\bibinfo {volume} {22}},\ \bibinfo {pages} {15}
  (\bibinfo {year} {2009})}\BibitemShut {NoStop}%
\bibitem [{\citenamefont {{Chen}}\ \emph {et~al.}(2022)\citenamefont {{Chen}},
  \citenamefont {{Kitamura}}, \citenamefont {{Barbieri}}, \citenamefont
  {{Nishiura}},\ and\ \citenamefont {{Furuichi}}}]{CHEN2022117304}%
  \BibitemOpen
  \bibfield  {author} {\bibinfo {author} {\bibfnamefont {J.}~\bibnamefont
  {{Chen}}}, \bibinfo {author} {\bibfnamefont {A.}~\bibnamefont {{Kitamura}}},
  \bibinfo {author} {\bibfnamefont {E.}~\bibnamefont {{Barbieri}}}, \bibinfo
  {author} {\bibfnamefont {D.}~\bibnamefont {{Nishiura}}},\ and\ \bibinfo
  {author} {\bibfnamefont {M.}~\bibnamefont {{Furuichi}}},\ }\bibfield  {title}
  {\bibinfo {title} {{Analyzing effects of microscopic material parameters on
  macroscopic mechanical responses in underwater mixing using discrete element
  method}},\ }\href
  {https://doi.org/https://doi.org/10.1016/j.powtec.2022.117304} {\bibfield
  {journal} {\bibinfo  {journal} {Powder Technology}\ }\textbf {\bibinfo
  {volume} {401}},\ \bibinfo {pages} {117304} (\bibinfo {year}
  {2022})}\BibitemShut {NoStop}%
\bibitem [{\citenamefont {{Hertz}}(1896)}]{hertz1896miscellaneous}%
  \BibitemOpen
  \bibfield  {author} {\bibinfo {author} {\bibfnamefont {H.}~\bibnamefont
  {{Hertz}}},\ }\href@noop {} {\emph {\bibinfo {title} {{Miscellaneous
  papers}}}}\ (\bibinfo  {publisher} {Macmillan},\ \bibinfo {year}
  {1896})\BibitemShut {NoStop}%
\bibitem [{\citenamefont {{Johnson}}\ \emph {et~al.}(1971)\citenamefont
  {{Johnson}}, \citenamefont {{Kendall}},\ and\ \citenamefont
  {{Roberts}}}]{1971RSPSA.324..301J}%
  \BibitemOpen
  \bibfield  {author} {\bibinfo {author} {\bibfnamefont {K.~L.}\ \bibnamefont
  {{Johnson}}}, \bibinfo {author} {\bibfnamefont {K.}~\bibnamefont
  {{Kendall}}},\ and\ \bibinfo {author} {\bibfnamefont {A.~D.}\ \bibnamefont
  {{Roberts}}},\ }\bibfield  {title} {\bibinfo {title} {{Surface Energy and the
  Contact of Elastic Solids}},\ }\href {https://doi.org/10.1098/rspa.1971.0141}
  {\bibfield  {journal} {\bibinfo  {journal} {Proceedings of the Royal Society
  of London Series A}\ }\textbf {\bibinfo {volume} {324}},\ \bibinfo {pages}
  {301} (\bibinfo {year} {1971})}\BibitemShut {NoStop}%
\bibitem [{\citenamefont {{Derjaguin}}\ \emph {et~al.}(1975)\citenamefont
  {{Derjaguin}}, \citenamefont {{Muller}},\ and\ \citenamefont
  {{Toporov}}}]{1975JCIS...53..314D}%
  \BibitemOpen
  \bibfield  {author} {\bibinfo {author} {\bibfnamefont {B.~V.}\ \bibnamefont
  {{Derjaguin}}}, \bibinfo {author} {\bibfnamefont {V.~M.}\ \bibnamefont
  {{Muller}}},\ and\ \bibinfo {author} {\bibfnamefont {Y.~P.}\ \bibnamefont
  {{Toporov}}},\ }\bibfield  {title} {\bibinfo {title} {{Effect of contact
  deformations on the adhesion of particles}},\ }\href
  {https://doi.org/10.1016/0021-9797(75)90018-1} {\bibfield  {journal}
  {\bibinfo  {journal} {Journal of Colloid and Interface Science}\ }\textbf
  {\bibinfo {volume} {53}},\ \bibinfo {pages} {314} (\bibinfo {year}
  {1975})}\BibitemShut {NoStop}%
\bibitem [{\citenamefont {{Chan}}\ and\ \citenamefont
  {{Tien}}(1973)}]{chan1973conductance}%
  \BibitemOpen
  \bibfield  {author} {\bibinfo {author} {\bibfnamefont {C.~K.}\ \bibnamefont
  {{Chan}}}\ and\ \bibinfo {author} {\bibfnamefont {C.~L.}\ \bibnamefont
  {{Tien}}},\ }\bibfield  {title} {\bibinfo {title} {{Conductance of packed
  spheres in vacuum}},\ }\href@noop {} {\bibfield  {journal} {\bibinfo
  {journal} {Journal of Heat Transfer}\ }\textbf {\bibinfo {volume} {95}},\
  \bibinfo {pages} {302} (\bibinfo {year} {1973})}\BibitemShut {NoStop}%
\bibitem [{\citenamefont {{Dominik}}\ and\ \citenamefont
  {{Tielens}}(1997)}]{1997ApJ...480..647D}%
  \BibitemOpen
  \bibfield  {author} {\bibinfo {author} {\bibfnamefont {C.}~\bibnamefont
  {{Dominik}}}\ and\ \bibinfo {author} {\bibfnamefont {A.~G.~G.~M.}\
  \bibnamefont {{Tielens}}},\ }\bibfield  {title} {\bibinfo {title} {{The
  Physics of Dust Coagulation and the Structure of Dust Aggregates in Space}},\
  }\href {https://doi.org/10.1086/303996} {\bibfield  {journal} {\bibinfo
  {journal} {\apj}\ }\textbf {\bibinfo {volume} {480}},\ \bibinfo {pages} {647}
  (\bibinfo {year} {1997})}\BibitemShut {NoStop}%
\bibitem [{\citenamefont {{Gusarov}}\ \emph {et~al.}(2003)\citenamefont
  {{Gusarov}}, \citenamefont {{Laoui}}, \citenamefont {{Froyen}},\ and\
  \citenamefont {{Titov}}}]{GUSAROV20031103}%
  \BibitemOpen
  \bibfield  {author} {\bibinfo {author} {\bibfnamefont {A.}~\bibnamefont
  {{Gusarov}}}, \bibinfo {author} {\bibfnamefont {T.}~\bibnamefont {{Laoui}}},
  \bibinfo {author} {\bibfnamefont {L.}~\bibnamefont {{Froyen}}},\ and\
  \bibinfo {author} {\bibfnamefont {V.}~\bibnamefont {{Titov}}},\ }\bibfield
  {title} {\bibinfo {title} {{Contact thermal conductivity of a powder bed in
  selective laser sintering}},\ }\href
  {https://doi.org/https://doi.org/10.1016/S0017-9310(02)00370-8} {\bibfield
  {journal} {\bibinfo  {journal} {International Journal of Heat and Mass
  Transfer}\ }\textbf {\bibinfo {volume} {46}},\ \bibinfo {pages} {1103}
  (\bibinfo {year} {2003})}\BibitemShut {NoStop}%
\bibitem [{\citenamefont {{Sakatani}}\ \emph {et~al.}(2017)\citenamefont
  {{Sakatani}}, \citenamefont {{Ogawa}}, \citenamefont {{Iijima}},
  \citenamefont {{Arakawa}}, \citenamefont {{Honda}},\ and\ \citenamefont
  {{Tanaka}}}]{2017AIPA....7a5310S}%
  \BibitemOpen
  \bibfield  {author} {\bibinfo {author} {\bibfnamefont {N.}~\bibnamefont
  {{Sakatani}}}, \bibinfo {author} {\bibfnamefont {K.}~\bibnamefont {{Ogawa}}},
  \bibinfo {author} {\bibfnamefont {Y.}~\bibnamefont {{Iijima}}}, \bibinfo
  {author} {\bibfnamefont {M.}~\bibnamefont {{Arakawa}}}, \bibinfo {author}
  {\bibfnamefont {R.}~\bibnamefont {{Honda}}},\ and\ \bibinfo {author}
  {\bibfnamefont {S.}~\bibnamefont {{Tanaka}}},\ }\bibfield  {title} {\bibinfo
  {title} {{Thermal conductivity model for powdered materials under vacuum
  based on experimental studies}},\ }\href {https://doi.org/10.1063/1.4975153}
  {\bibfield  {journal} {\bibinfo  {journal} {AIP Advances}\ }\textbf {\bibinfo
  {volume} {7}},\ \bibinfo {eid} {015310} (\bibinfo {year} {2017})}\BibitemShut
  {NoStop}%
\bibitem [{\citenamefont {{Arakawa}}\ \emph {et~al.}(2017)\citenamefont
  {{Arakawa}}, \citenamefont {{Tanaka}}, \citenamefont {{Kataoka}},\ and\
  \citenamefont {{Nakamoto}}}]{2017A&A...608L...7A}%
  \BibitemOpen
  \bibfield  {author} {\bibinfo {author} {\bibfnamefont {S.}~\bibnamefont
  {{Arakawa}}}, \bibinfo {author} {\bibfnamefont {H.}~\bibnamefont {{Tanaka}}},
  \bibinfo {author} {\bibfnamefont {A.}~\bibnamefont {{Kataoka}}},\ and\
  \bibinfo {author} {\bibfnamefont {T.}~\bibnamefont {{Nakamoto}}},\ }\bibfield
   {title} {\bibinfo {title} {{Thermal conductivity of porous aggregates}},\
  }\href {https://doi.org/10.1051/0004-6361/201732182} {\bibfield  {journal}
  {\bibinfo  {journal} {Astronomy \& Astrophysics}\ }\textbf {\bibinfo {volume}
  {608}},\ \bibinfo {eid} {L7} (\bibinfo {year} {2017})}\BibitemShut {NoStop}%
\bibitem [{\citenamefont {{Ben-Nun}}\ \emph {et~al.}(2010)\citenamefont
  {{Ben-Nun}}, \citenamefont {{Einav}},\ and\ \citenamefont
  {{Tordesillas}}}]{2010PhRvL.104j8001B}%
  \BibitemOpen
  \bibfield  {author} {\bibinfo {author} {\bibfnamefont {O.}~\bibnamefont
  {{Ben-Nun}}}, \bibinfo {author} {\bibfnamefont {I.}~\bibnamefont {{Einav}}},\
  and\ \bibinfo {author} {\bibfnamefont {A.}~\bibnamefont {{Tordesillas}}},\
  }\bibfield  {title} {\bibinfo {title} {{Force Attractor in Confined
  Comminution of Granular Materials}},\ }\href
  {https://doi.org/10.1103/PhysRevLett.104.108001} {\bibfield  {journal}
  {\bibinfo  {journal} {\prl}\ }\textbf {\bibinfo {volume} {104}},\ \bibinfo
  {eid} {108001} (\bibinfo {year} {2010})}\BibitemShut {NoStop}%
\bibitem [{\citenamefont {{Schr{\"a}pler}}\ \emph {et~al.}(2015)\citenamefont
  {{Schr{\"a}pler}}, \citenamefont {{Blum}}, \citenamefont {{von Borstel}},\
  and\ \citenamefont {{G{\"u}ttler}}}]{2015Icar..257...33S}%
  \BibitemOpen
  \bibfield  {author} {\bibinfo {author} {\bibfnamefont {R.}~\bibnamefont
  {{Schr{\"a}pler}}}, \bibinfo {author} {\bibfnamefont {J.}~\bibnamefont
  {{Blum}}}, \bibinfo {author} {\bibfnamefont {I.}~\bibnamefont {{von
  Borstel}}},\ and\ \bibinfo {author} {\bibfnamefont {C.}~\bibnamefont
  {{G{\"u}ttler}}},\ }\bibfield  {title} {\bibinfo {title} {{The stratification
  of regolith on celestial objects}},\ }\href
  {https://doi.org/10.1016/j.icarus.2015.04.033} {\bibfield  {journal}
  {\bibinfo  {journal} {Icarus}\ }\textbf {\bibinfo {volume} {257}},\ \bibinfo
  {pages} {33} (\bibinfo {year} {2015})}\BibitemShut {NoStop}%
\bibitem [{\citenamefont {{Okubo}}\ and\ \citenamefont
  {{Katsuragi}}(2022)}]{2022A&A...664A.147O}%
  \BibitemOpen
  \bibfield  {author} {\bibinfo {author} {\bibfnamefont {F.}~\bibnamefont
  {{Okubo}}}\ and\ \bibinfo {author} {\bibfnamefont {H.}~\bibnamefont
  {{Katsuragi}}},\ }\bibfield  {title} {\bibinfo {title} {{Impact drag force
  exerted on a projectile penetrating into a hierarchical granular bed}},\
  }\href {https://doi.org/10.1051/0004-6361/202243787} {\bibfield  {journal}
  {\bibinfo  {journal} {Astronomy \& Astrophysics}\ }\textbf {\bibinfo {volume}
  {664}},\ \bibinfo {eid} {A147} (\bibinfo {year} {2022})}\BibitemShut
  {NoStop}%
\bibitem [{\citenamefont {{Mueth}}\ \emph {et~al.}(1998)\citenamefont
  {{Mueth}}, \citenamefont {{Jaeger}},\ and\ \citenamefont
  {{Nagel}}}]{1998PhRvE..57.3164M}%
  \BibitemOpen
  \bibfield  {author} {\bibinfo {author} {\bibfnamefont {D.~M.}\ \bibnamefont
  {{Mueth}}}, \bibinfo {author} {\bibfnamefont {H.~M.}\ \bibnamefont
  {{Jaeger}}},\ and\ \bibinfo {author} {\bibfnamefont {S.~R.}\ \bibnamefont
  {{Nagel}}},\ }\bibfield  {title} {\bibinfo {title} {{Force distribution in a
  granular medium}},\ }\href {https://doi.org/10.1103/PhysRevE.57.3164}
  {\bibfield  {journal} {\bibinfo  {journal} {\pre}\ }\textbf {\bibinfo
  {volume} {57}},\ \bibinfo {pages} {3164} (\bibinfo {year}
  {1998})}\BibitemShut {NoStop}%
\bibitem [{\citenamefont {{O'Hern}}\ \emph {et~al.}(2001)\citenamefont
  {{O'Hern}}, \citenamefont {{Langer}}, \citenamefont {{Liu}},\ and\
  \citenamefont {{Nagel}}}]{2001PhRvL..86..111O}%
  \BibitemOpen
  \bibfield  {author} {\bibinfo {author} {\bibfnamefont {C.~S.}\ \bibnamefont
  {{O'Hern}}}, \bibinfo {author} {\bibfnamefont {S.~A.}\ \bibnamefont
  {{Langer}}}, \bibinfo {author} {\bibfnamefont {A.~J.}\ \bibnamefont
  {{Liu}}},\ and\ \bibinfo {author} {\bibfnamefont {S.~R.}\ \bibnamefont
  {{Nagel}}},\ }\bibfield  {title} {\bibinfo {title} {{Force Distributions near
  Jamming and Glass Transitions}},\ }\href
  {https://doi.org/10.1103/PhysRevLett.86.111} {\bibfield  {journal} {\bibinfo
  {journal} {\prl}\ }\textbf {\bibinfo {volume} {86}},\ \bibinfo {pages} {111}
  (\bibinfo {year} {2001})}\BibitemShut {NoStop}%
\bibitem [{\citenamefont {{Majmudar}}\ and\ \citenamefont
  {{Behringer}}(2005)}]{2005Natur.435.1079M}%
  \BibitemOpen
  \bibfield  {author} {\bibinfo {author} {\bibfnamefont {T.~S.}\ \bibnamefont
  {{Majmudar}}}\ and\ \bibinfo {author} {\bibfnamefont {R.~P.}\ \bibnamefont
  {{Behringer}}},\ }\bibfield  {title} {\bibinfo {title} {{Contact force
  measurements and stress-induced anisotropy in granular materials}},\ }\href
  {https://doi.org/10.1038/nature03805} {\bibfield  {journal} {\bibinfo
  {journal} {Nature}\ }\textbf {\bibinfo {volume} {435}},\ \bibinfo {pages}
  {1079} (\bibinfo {year} {2005})}\BibitemShut {NoStop}%
\bibitem [{\citenamefont {{G{\"u}ttler}}\ \emph {et~al.}(2009)\citenamefont
  {{G{\"u}ttler}}, \citenamefont {{Krause}}, \citenamefont {{Geretshauser}},
  \citenamefont {{Speith}},\ and\ \citenamefont
  {{Blum}}}]{2009ApJ...701..130G}%
  \BibitemOpen
  \bibfield  {author} {\bibinfo {author} {\bibfnamefont {C.}~\bibnamefont
  {{G{\"u}ttler}}}, \bibinfo {author} {\bibfnamefont {M.}~\bibnamefont
  {{Krause}}}, \bibinfo {author} {\bibfnamefont {R.~J.}\ \bibnamefont
  {{Geretshauser}}}, \bibinfo {author} {\bibfnamefont {R.}~\bibnamefont
  {{Speith}}},\ and\ \bibinfo {author} {\bibfnamefont {J.}~\bibnamefont
  {{Blum}}},\ }\bibfield  {title} {\bibinfo {title} {{The Physics of
  Protoplanetesimal Dust Agglomerates. IV. Toward a Dynamical Collision
  Model}},\ }\href {https://doi.org/10.1088/0004-637X/701/1/130} {\bibfield
  {journal} {\bibinfo  {journal} {\apj}\ }\textbf {\bibinfo {volume} {701}},\
  \bibinfo {pages} {130} (\bibinfo {year} {2009})}\BibitemShut {NoStop}%
\bibitem [{\citenamefont {{Kataoka}}\ \emph
  {et~al.}(2013{\natexlab{a}})\citenamefont {{Kataoka}}, \citenamefont
  {{Tanaka}}, \citenamefont {{Okuzumi}},\ and\ \citenamefont
  {{Wada}}}]{2013A&A...554A...4K}%
  \BibitemOpen
  \bibfield  {author} {\bibinfo {author} {\bibfnamefont {A.}~\bibnamefont
  {{Kataoka}}}, \bibinfo {author} {\bibfnamefont {H.}~\bibnamefont {{Tanaka}}},
  \bibinfo {author} {\bibfnamefont {S.}~\bibnamefont {{Okuzumi}}},\ and\
  \bibinfo {author} {\bibfnamefont {K.}~\bibnamefont {{Wada}}},\ }\bibfield
  {title} {\bibinfo {title} {{Static compression of porous dust aggregates}},\
  }\href {https://doi.org/10.1051/0004-6361/201321325} {\bibfield  {journal}
  {\bibinfo  {journal} {Astronomy \& Astrophysics}\ }\textbf {\bibinfo {volume}
  {554}},\ \bibinfo {eid} {A4} (\bibinfo {year}
  {2013}{\natexlab{a}})}\BibitemShut {NoStop}%
\bibitem [{\citenamefont {{Omura}}\ and\ \citenamefont
  {{Nakamura}}(2017)}]{2017P&SS..149...14O}%
  \BibitemOpen
  \bibfield  {author} {\bibinfo {author} {\bibfnamefont {T.}~\bibnamefont
  {{Omura}}}\ and\ \bibinfo {author} {\bibfnamefont {A.~M.}\ \bibnamefont
  {{Nakamura}}},\ }\bibfield  {title} {\bibinfo {title} {{Experimental study on
  compression property of regolith analogues}},\ }\href
  {https://doi.org/10.1016/j.pss.2017.08.003} {\bibfield  {journal} {\bibinfo
  {journal} {Planetary and Space Science}\ }\textbf {\bibinfo {volume} {149}},\
  \bibinfo {pages} {14} (\bibinfo {year} {2017})}\BibitemShut {NoStop}%
\bibitem [{\citenamefont {{Okuzumi}}\ \emph {et~al.}(2012)\citenamefont
  {{Okuzumi}}, \citenamefont {{Tanaka}}, \citenamefont {{Kobayashi}},\ and\
  \citenamefont {{Wada}}}]{2012ApJ...752..106O}%
  \BibitemOpen
  \bibfield  {author} {\bibinfo {author} {\bibfnamefont {S.}~\bibnamefont
  {{Okuzumi}}}, \bibinfo {author} {\bibfnamefont {H.}~\bibnamefont {{Tanaka}}},
  \bibinfo {author} {\bibfnamefont {H.}~\bibnamefont {{Kobayashi}}},\ and\
  \bibinfo {author} {\bibfnamefont {K.}~\bibnamefont {{Wada}}},\ }\bibfield
  {title} {\bibinfo {title} {{Rapid Coagulation of Porous Dust Aggregates
  outside the Snow Line: A Pathway to Successful Icy Planetesimal Formation}},\
  }\href {https://doi.org/10.1088/0004-637X/752/2/106} {\bibfield  {journal}
  {\bibinfo  {journal} {\apj}\ }\textbf {\bibinfo {volume} {752}},\ \bibinfo
  {eid} {106} (\bibinfo {year} {2012})}\BibitemShut {NoStop}%
\bibitem [{\citenamefont {{Kataoka}}\ \emph
  {et~al.}(2013{\natexlab{b}})\citenamefont {{Kataoka}}, \citenamefont
  {{Tanaka}}, \citenamefont {{Okuzumi}},\ and\ \citenamefont
  {{Wada}}}]{2013A&A...557L...4K}%
  \BibitemOpen
  \bibfield  {author} {\bibinfo {author} {\bibfnamefont {A.}~\bibnamefont
  {{Kataoka}}}, \bibinfo {author} {\bibfnamefont {H.}~\bibnamefont {{Tanaka}}},
  \bibinfo {author} {\bibfnamefont {S.}~\bibnamefont {{Okuzumi}}},\ and\
  \bibinfo {author} {\bibfnamefont {K.}~\bibnamefont {{Wada}}},\ }\bibfield
  {title} {\bibinfo {title} {{Fluffy dust forms icy planetesimals by static
  compression}},\ }\href {https://doi.org/10.1051/0004-6361/201322151}
  {\bibfield  {journal} {\bibinfo  {journal} {Astronomy \& Astrophysics}\
  }\textbf {\bibinfo {volume} {557}},\ \bibinfo {eid} {L4} (\bibinfo {year}
  {2013}{\natexlab{b}})}\BibitemShut {NoStop}%
\bibitem [{\citenamefont {{Okada}}\ \emph {et~al.}(2020)\citenamefont
  {{Okada}}, \citenamefont {{Fukuhara}}, \citenamefont {{Tanaka}},
  \citenamefont {{Taguchi}}, \citenamefont {{Arai}}, \citenamefont {{Senshu}},
  \citenamefont {{Sakatani}}, \citenamefont {{Shimaki}}, \citenamefont
  {{Demura}}, \citenamefont {{Ogawa}}, \citenamefont {{Suko}}, \citenamefont
  {{Sekiguchi}}, \citenamefont {{Kouyama}}, \citenamefont {{Takita}},
  \citenamefont {{Matsunaga}}, \citenamefont {{Imamura}}, \citenamefont
  {{Wada}}, \citenamefont {{Hasegawa}}, \citenamefont {{Helbert}},
  \citenamefont {{M{\"u}ller}}, \citenamefont {{Hagermann}}, \citenamefont
  {{Biele}}, \citenamefont {{Grott}}, \citenamefont {{Hamm}}, \citenamefont
  {{Delbo}}, \citenamefont {{Hirata}}, \citenamefont {{Hirata}}, \citenamefont
  {{Yamamoto}}, \citenamefont {{Sugita}}, \citenamefont {{Namiki}},
  \citenamefont {{Kitazato}}, \citenamefont {{Arakawa}}, \citenamefont
  {{Tachibana}}, \citenamefont {{Ikeda}}, \citenamefont {{Ishiguro}},
  \citenamefont {{Wada}}, \citenamefont {{Honda}}, \citenamefont {{Honda}},
  \citenamefont {{Ishihara}}, \citenamefont {{Matsumoto}}, \citenamefont
  {{Matsuoka}}, \citenamefont {{Michikami}}, \citenamefont {{Miura}},
  \citenamefont {{Morota}}, \citenamefont {{Noda}}, \citenamefont {{Noguchi}},
  \citenamefont {{Ogawa}}, \citenamefont {{Shirai}}, \citenamefont {{Tatsumi}},
  \citenamefont {{Yabuta}}, \citenamefont {{Yokota}}, \citenamefont {{Yamada}},
  \citenamefont {{Abe}}, \citenamefont {{Hayakawa}}, \citenamefont {{Iwata}},
  \citenamefont {{Ozaki}}, \citenamefont {{Yano}}, \citenamefont {{Hosoda}},
  \citenamefont {{Mori}}, \citenamefont {{Sawada}}, \citenamefont {{Shimada}},
  \citenamefont {{Takeuchi}}, \citenamefont {{Tsukizaki}}, \citenamefont
  {{Fujii}}, \citenamefont {{Hirose}}, \citenamefont {{Kikuchi}}, \citenamefont
  {{Mimasu}}, \citenamefont {{Ogawa}}, \citenamefont {{Ono}}, \citenamefont
  {{Takahashi}}, \citenamefont {{Takei}}, \citenamefont {{Yamaguchi}},
  \citenamefont {{Yoshikawa}}, \citenamefont {{Terui}}, \citenamefont
  {{Saiki}}, \citenamefont {{Nakazawa}}, \citenamefont {{Yoshikawa}},
  \citenamefont {{Watanabe}},\ and\ \citenamefont
  {{Tsuda}}}]{2020Natur.579..518O}%
  \BibitemOpen
  \bibfield  {author} {\bibinfo {author} {\bibfnamefont {T.}~\bibnamefont
  {{Okada}}}, \bibinfo {author} {\bibfnamefont {T.}~\bibnamefont {{Fukuhara}}},
  \bibinfo {author} {\bibfnamefont {S.}~\bibnamefont {{Tanaka}}}, \bibinfo
  {author} {\bibfnamefont {M.}~\bibnamefont {{Taguchi}}}, \bibinfo {author}
  {\bibfnamefont {T.}~\bibnamefont {{Arai}}}, \bibinfo {author} {\bibfnamefont
  {H.}~\bibnamefont {{Senshu}}}, \bibinfo {author} {\bibfnamefont
  {N.}~\bibnamefont {{Sakatani}}}, \bibinfo {author} {\bibfnamefont
  {Y.}~\bibnamefont {{Shimaki}}}, \bibinfo {author} {\bibfnamefont
  {H.}~\bibnamefont {{Demura}}}, \bibinfo {author} {\bibfnamefont
  {Y.}~\bibnamefont {{Ogawa}}}, \bibinfo {author} {\bibfnamefont
  {K.}~\bibnamefont {{Suko}}}, \bibinfo {author} {\bibfnamefont
  {T.}~\bibnamefont {{Sekiguchi}}}, \bibinfo {author} {\bibfnamefont
  {T.}~\bibnamefont {{Kouyama}}}, \bibinfo {author} {\bibfnamefont
  {J.}~\bibnamefont {{Takita}}}, \bibinfo {author} {\bibfnamefont
  {T.}~\bibnamefont {{Matsunaga}}}, \bibinfo {author} {\bibfnamefont
  {T.}~\bibnamefont {{Imamura}}}, \bibinfo {author} {\bibfnamefont
  {T.}~\bibnamefont {{Wada}}}, \bibinfo {author} {\bibfnamefont
  {S.}~\bibnamefont {{Hasegawa}}}, \bibinfo {author} {\bibfnamefont
  {J.}~\bibnamefont {{Helbert}}}, \bibinfo {author} {\bibfnamefont {T.~G.}\
  \bibnamefont {{M{\"u}ller}}}, \bibinfo {author} {\bibfnamefont
  {A.}~\bibnamefont {{Hagermann}}}, \bibinfo {author} {\bibfnamefont
  {J.}~\bibnamefont {{Biele}}}, \bibinfo {author} {\bibfnamefont
  {M.}~\bibnamefont {{Grott}}}, \bibinfo {author} {\bibfnamefont
  {M.}~\bibnamefont {{Hamm}}}, \bibinfo {author} {\bibfnamefont
  {M.}~\bibnamefont {{Delbo}}}, \bibinfo {author} {\bibfnamefont
  {N.}~\bibnamefont {{Hirata}}}, \bibinfo {author} {\bibfnamefont
  {N.}~\bibnamefont {{Hirata}}}, \bibinfo {author} {\bibfnamefont
  {Y.}~\bibnamefont {{Yamamoto}}}, \bibinfo {author} {\bibfnamefont
  {S.}~\bibnamefont {{Sugita}}}, \bibinfo {author} {\bibfnamefont
  {N.}~\bibnamefont {{Namiki}}}, \bibinfo {author} {\bibfnamefont
  {K.}~\bibnamefont {{Kitazato}}}, \bibinfo {author} {\bibfnamefont
  {M.}~\bibnamefont {{Arakawa}}}, \bibinfo {author} {\bibfnamefont
  {S.}~\bibnamefont {{Tachibana}}}, \bibinfo {author} {\bibfnamefont
  {H.}~\bibnamefont {{Ikeda}}}, \bibinfo {author} {\bibfnamefont
  {M.}~\bibnamefont {{Ishiguro}}}, \bibinfo {author} {\bibfnamefont
  {K.}~\bibnamefont {{Wada}}}, \bibinfo {author} {\bibfnamefont
  {C.}~\bibnamefont {{Honda}}}, \bibinfo {author} {\bibfnamefont
  {R.}~\bibnamefont {{Honda}}}, \bibinfo {author} {\bibfnamefont
  {Y.}~\bibnamefont {{Ishihara}}}, \bibinfo {author} {\bibfnamefont
  {K.}~\bibnamefont {{Matsumoto}}}, \bibinfo {author} {\bibfnamefont
  {M.}~\bibnamefont {{Matsuoka}}}, \bibinfo {author} {\bibfnamefont
  {T.}~\bibnamefont {{Michikami}}}, \bibinfo {author} {\bibfnamefont
  {A.}~\bibnamefont {{Miura}}}, \bibinfo {author} {\bibfnamefont
  {T.}~\bibnamefont {{Morota}}}, \bibinfo {author} {\bibfnamefont
  {H.}~\bibnamefont {{Noda}}}, \bibinfo {author} {\bibfnamefont
  {R.}~\bibnamefont {{Noguchi}}}, \bibinfo {author} {\bibfnamefont
  {K.}~\bibnamefont {{Ogawa}}}, \bibinfo {author} {\bibfnamefont
  {K.}~\bibnamefont {{Shirai}}}, \bibinfo {author} {\bibfnamefont
  {E.}~\bibnamefont {{Tatsumi}}}, \bibinfo {author} {\bibfnamefont
  {H.}~\bibnamefont {{Yabuta}}}, \bibinfo {author} {\bibfnamefont
  {Y.}~\bibnamefont {{Yokota}}}, \bibinfo {author} {\bibfnamefont
  {M.}~\bibnamefont {{Yamada}}}, \bibinfo {author} {\bibfnamefont
  {M.}~\bibnamefont {{Abe}}}, \bibinfo {author} {\bibfnamefont
  {M.}~\bibnamefont {{Hayakawa}}}, \bibinfo {author} {\bibfnamefont
  {T.}~\bibnamefont {{Iwata}}}, \bibinfo {author} {\bibfnamefont
  {M.}~\bibnamefont {{Ozaki}}}, \bibinfo {author} {\bibfnamefont
  {H.}~\bibnamefont {{Yano}}}, \bibinfo {author} {\bibfnamefont
  {S.}~\bibnamefont {{Hosoda}}}, \bibinfo {author} {\bibfnamefont
  {O.}~\bibnamefont {{Mori}}}, \bibinfo {author} {\bibfnamefont
  {H.}~\bibnamefont {{Sawada}}}, \bibinfo {author} {\bibfnamefont
  {T.}~\bibnamefont {{Shimada}}}, \bibinfo {author} {\bibfnamefont
  {H.}~\bibnamefont {{Takeuchi}}}, \bibinfo {author} {\bibfnamefont
  {R.}~\bibnamefont {{Tsukizaki}}}, \bibinfo {author} {\bibfnamefont
  {A.}~\bibnamefont {{Fujii}}}, \bibinfo {author} {\bibfnamefont
  {C.}~\bibnamefont {{Hirose}}}, \bibinfo {author} {\bibfnamefont
  {S.}~\bibnamefont {{Kikuchi}}}, \bibinfo {author} {\bibfnamefont
  {Y.}~\bibnamefont {{Mimasu}}}, \bibinfo {author} {\bibfnamefont
  {N.}~\bibnamefont {{Ogawa}}}, \bibinfo {author} {\bibfnamefont
  {G.}~\bibnamefont {{Ono}}}, \bibinfo {author} {\bibfnamefont
  {T.}~\bibnamefont {{Takahashi}}}, \bibinfo {author} {\bibfnamefont
  {Y.}~\bibnamefont {{Takei}}}, \bibinfo {author} {\bibfnamefont
  {T.}~\bibnamefont {{Yamaguchi}}}, \bibinfo {author} {\bibfnamefont
  {K.}~\bibnamefont {{Yoshikawa}}}, \bibinfo {author} {\bibfnamefont
  {F.}~\bibnamefont {{Terui}}}, \bibinfo {author} {\bibfnamefont
  {T.}~\bibnamefont {{Saiki}}}, \bibinfo {author} {\bibfnamefont
  {S.}~\bibnamefont {{Nakazawa}}}, \bibinfo {author} {\bibfnamefont
  {M.}~\bibnamefont {{Yoshikawa}}}, \bibinfo {author} {\bibfnamefont
  {S.}~\bibnamefont {{Watanabe}}},\ and\ \bibinfo {author} {\bibfnamefont
  {Y.}~\bibnamefont {{Tsuda}}},\ }\bibfield  {title} {\bibinfo {title} {{Highly
  porous nature of a primitive asteroid revealed by thermal imaging}},\ }\href
  {https://doi.org/10.1038/s41586-020-2102-6} {\bibfield  {journal} {\bibinfo
  {journal} {Nature}\ }\textbf {\bibinfo {volume} {579}},\ \bibinfo {pages}
  {518} (\bibinfo {year} {2020})}\BibitemShut {NoStop}%
\bibitem [{\citenamefont {{Kobayashi}}\ and\ \citenamefont
  {{Tanaka}}(2021)}]{2021ApJ...922...16K}%
  \BibitemOpen
  \bibfield  {author} {\bibinfo {author} {\bibfnamefont {H.}~\bibnamefont
  {{Kobayashi}}}\ and\ \bibinfo {author} {\bibfnamefont {H.}~\bibnamefont
  {{Tanaka}}},\ }\bibfield  {title} {\bibinfo {title} {{Rapid Formation of
  Gas-giant Planets via Collisional Coagulation from Dust Grains to Planetary
  Cores}},\ }\href {https://doi.org/10.3847/1538-4357/ac289c} {\bibfield
  {journal} {\bibinfo  {journal} {\apj}\ }\textbf {\bibinfo {volume} {922}},\
  \bibinfo {eid} {16} (\bibinfo {year} {2021})}\BibitemShut {NoStop}%
\bibitem [{\citenamefont {{Tazaki}}\ \emph {et~al.}(2023)\citenamefont
  {{Tazaki}}, \citenamefont {{Ginski}},\ and\ \citenamefont
  {{Dominik}}}]{2023ApJ...944L..43T}%
  \BibitemOpen
  \bibfield  {author} {\bibinfo {author} {\bibfnamefont {R.}~\bibnamefont
  {{Tazaki}}}, \bibinfo {author} {\bibfnamefont {C.}~\bibnamefont {{Ginski}}},\
  and\ \bibinfo {author} {\bibfnamefont {C.}~\bibnamefont {{Dominik}}},\
  }\bibfield  {title} {\bibinfo {title} {{Fractal Aggregates of Submicron-sized
  Grains in the Young Planet-forming Disk around IM Lup}},\ }\href
  {https://doi.org/10.3847/2041-8213/acb824} {\bibfield  {journal} {\bibinfo
  {journal} {The Astrophysical Journal Letters}\ }\textbf {\bibinfo {volume}
  {944}},\ \bibinfo {eid} {L43} (\bibinfo {year} {2023})}\BibitemShut {NoStop}%
\bibitem [{\citenamefont {{Blum}}\ \emph {et~al.}(2000)\citenamefont {{Blum}},
  \citenamefont {{Wurm}}, \citenamefont {{Kempf}}, \citenamefont {{Poppe}},
  \citenamefont {{Klahr}}, \citenamefont {{Kozasa}}, \citenamefont {{Rott}},
  \citenamefont {{Henning}}, \citenamefont {{Dorschner}}, \citenamefont
  {{Schr{\"a}pler}}, \citenamefont {{Keller}}, \citenamefont {{Markiewicz}},
  \citenamefont {{Mann}}, \citenamefont {{Gustafson}}, \citenamefont
  {{Giovane}}, \citenamefont {{Neuhaus}}, \citenamefont {{Fechtig}},
  \citenamefont {{Gr{\"u}n}}, \citenamefont {{Feuerbacher}}, \citenamefont
  {{Kochan}}, \citenamefont {{Ratke}}, \citenamefont {{El Goresy}},
  \citenamefont {{Morfill}}, \citenamefont {{Weidenschilling}}, \citenamefont
  {{Schwehm}}, \citenamefont {{Metzler}},\ and\ \citenamefont
  {{Ip}}}]{2000PhRvL..85.2426B}%
  \BibitemOpen
  \bibfield  {author} {\bibinfo {author} {\bibfnamefont {J.}~\bibnamefont
  {{Blum}}}, \bibinfo {author} {\bibfnamefont {G.}~\bibnamefont {{Wurm}}},
  \bibinfo {author} {\bibfnamefont {S.}~\bibnamefont {{Kempf}}}, \bibinfo
  {author} {\bibfnamefont {T.}~\bibnamefont {{Poppe}}}, \bibinfo {author}
  {\bibfnamefont {H.}~\bibnamefont {{Klahr}}}, \bibinfo {author} {\bibfnamefont
  {T.}~\bibnamefont {{Kozasa}}}, \bibinfo {author} {\bibfnamefont
  {M.}~\bibnamefont {{Rott}}}, \bibinfo {author} {\bibfnamefont
  {T.}~\bibnamefont {{Henning}}}, \bibinfo {author} {\bibfnamefont
  {J.}~\bibnamefont {{Dorschner}}}, \bibinfo {author} {\bibfnamefont
  {R.}~\bibnamefont {{Schr{\"a}pler}}}, \bibinfo {author} {\bibfnamefont
  {H.~U.}\ \bibnamefont {{Keller}}}, \bibinfo {author} {\bibfnamefont {W.~J.}\
  \bibnamefont {{Markiewicz}}}, \bibinfo {author} {\bibfnamefont
  {I.}~\bibnamefont {{Mann}}}, \bibinfo {author} {\bibfnamefont {B.~A.}\
  \bibnamefont {{Gustafson}}}, \bibinfo {author} {\bibfnamefont
  {F.}~\bibnamefont {{Giovane}}}, \bibinfo {author} {\bibfnamefont
  {D.}~\bibnamefont {{Neuhaus}}}, \bibinfo {author} {\bibfnamefont
  {H.}~\bibnamefont {{Fechtig}}}, \bibinfo {author} {\bibfnamefont
  {E.}~\bibnamefont {{Gr{\"u}n}}}, \bibinfo {author} {\bibfnamefont
  {B.}~\bibnamefont {{Feuerbacher}}}, \bibinfo {author} {\bibfnamefont
  {H.}~\bibnamefont {{Kochan}}}, \bibinfo {author} {\bibfnamefont
  {L.}~\bibnamefont {{Ratke}}}, \bibinfo {author} {\bibfnamefont
  {A.}~\bibnamefont {{El Goresy}}}, \bibinfo {author} {\bibfnamefont
  {G.}~\bibnamefont {{Morfill}}}, \bibinfo {author} {\bibfnamefont {S.~J.}\
  \bibnamefont {{Weidenschilling}}}, \bibinfo {author} {\bibfnamefont
  {G.}~\bibnamefont {{Schwehm}}}, \bibinfo {author} {\bibfnamefont
  {K.}~\bibnamefont {{Metzler}}},\ and\ \bibinfo {author} {\bibfnamefont
  {W.~H.}\ \bibnamefont {{Ip}}},\ }\bibfield  {title} {\bibinfo {title}
  {{Growth and Form of Planetary Seedlings: Results from a Microgravity
  Aggregation Experiment}},\ }\href
  {https://doi.org/10.1103/PhysRevLett.85.2426} {\bibfield  {journal} {\bibinfo
   {journal} {\prl}\ }\textbf {\bibinfo {volume} {85}},\ \bibinfo {pages}
  {2426} (\bibinfo {year} {2000})}\BibitemShut {NoStop}%
\bibitem [{\citenamefont {{Paszun}}\ and\ \citenamefont
  {{Dominik}}(2006)}]{2006Icar..182..274P}%
  \BibitemOpen
  \bibfield  {author} {\bibinfo {author} {\bibfnamefont {D.}~\bibnamefont
  {{Paszun}}}\ and\ \bibinfo {author} {\bibfnamefont {C.}~\bibnamefont
  {{Dominik}}},\ }\bibfield  {title} {\bibinfo {title} {{The influence of grain
  rotation on the structure of dust aggregates}},\ }\href
  {https://doi.org/10.1016/j.icarus.2005.12.018} {\bibfield  {journal}
  {\bibinfo  {journal} {Icarus}\ }\textbf {\bibinfo {volume} {182}},\ \bibinfo
  {pages} {274} (\bibinfo {year} {2006})}\BibitemShut {NoStop}%
\bibitem [{\citenamefont {{Suyama}}\ \emph {et~al.}(2008)\citenamefont
  {{Suyama}}, \citenamefont {{Wada}},\ and\ \citenamefont
  {{Tanaka}}}]{2008ApJ...684.1310S}%
  \BibitemOpen
  \bibfield  {author} {\bibinfo {author} {\bibfnamefont {T.}~\bibnamefont
  {{Suyama}}}, \bibinfo {author} {\bibfnamefont {K.}~\bibnamefont {{Wada}}},\
  and\ \bibinfo {author} {\bibfnamefont {H.}~\bibnamefont {{Tanaka}}},\
  }\bibfield  {title} {\bibinfo {title} {{Numerical Simulation of Density
  Evolution of Dust Aggregates in Protoplanetary Disks. I. Head-on
  Collisions}},\ }\href {https://doi.org/10.1086/590143} {\bibfield  {journal}
  {\bibinfo  {journal} {\apj}\ }\textbf {\bibinfo {volume} {684}},\ \bibinfo
  {pages} {1310} (\bibinfo {year} {2008})}\BibitemShut {NoStop}%
\bibitem [{\citenamefont {{Meakin}}(1999)}]{WOS:000081472000002}%
  \BibitemOpen
  \bibfield  {author} {\bibinfo {author} {\bibfnamefont {P.}~\bibnamefont
  {{Meakin}}},\ }\bibfield  {title} {\bibinfo {title} {{A historical
  introduction to computer models for fractal aggregates}},\ }\href
  {https://doi.org/10.1023/A:1008731904082} {\bibfield  {journal} {\bibinfo
  {journal} {Journal of Sol-Gel Science and Technology}\ }\textbf {\bibinfo
  {volume} {15}},\ \bibinfo {pages} {97} (\bibinfo {year} {1999})}\BibitemShut
  {NoStop}%
\bibitem [{\citenamefont {{Blum}}\ and\ \citenamefont
  {{Wurm}}(2000)}]{2000Icar..143..138B}%
  \BibitemOpen
  \bibfield  {author} {\bibinfo {author} {\bibfnamefont {J.}~\bibnamefont
  {{Blum}}}\ and\ \bibinfo {author} {\bibfnamefont {G.}~\bibnamefont
  {{Wurm}}},\ }\bibfield  {title} {\bibinfo {title} {{Experiments on Sticking,
  Restructuring, and Fragmentation of Preplanetary Dust Aggregates}},\ }\href
  {https://doi.org/10.1006/icar.1999.6234} {\bibfield  {journal} {\bibinfo
  {journal} {Icarus}\ }\textbf {\bibinfo {volume} {143}},\ \bibinfo {pages}
  {138} (\bibinfo {year} {2000})}\BibitemShut {NoStop}%
\bibitem [{\citenamefont {{Wada}}\ \emph {et~al.}(2007)\citenamefont {{Wada}},
  \citenamefont {{Tanaka}}, \citenamefont {{Suyama}}, \citenamefont
  {{Kimura}},\ and\ \citenamefont {{Yamamoto}}}]{2007ApJ...661..320W}%
  \BibitemOpen
  \bibfield  {author} {\bibinfo {author} {\bibfnamefont {K.}~\bibnamefont
  {{Wada}}}, \bibinfo {author} {\bibfnamefont {H.}~\bibnamefont {{Tanaka}}},
  \bibinfo {author} {\bibfnamefont {T.}~\bibnamefont {{Suyama}}}, \bibinfo
  {author} {\bibfnamefont {H.}~\bibnamefont {{Kimura}}},\ and\ \bibinfo
  {author} {\bibfnamefont {T.}~\bibnamefont {{Yamamoto}}},\ }\bibfield  {title}
  {\bibinfo {title} {{Numerical Simulation of Dust Aggregate Collisions. I.
  Compression and Disruption of Two-Dimensional Aggregates}},\ }\href
  {https://doi.org/10.1086/514332} {\bibfield  {journal} {\bibinfo  {journal}
  {\apj}\ }\textbf {\bibinfo {volume} {661}},\ \bibinfo {pages} {320} (\bibinfo
  {year} {2007})}\BibitemShut {NoStop}%
\bibitem [{\citenamefont {{Tatsuuma}}\ \emph {et~al.}(2019)\citenamefont
  {{Tatsuuma}}, \citenamefont {{Kataoka}},\ and\ \citenamefont
  {{Tanaka}}}]{2019ApJ...874..159T}%
  \BibitemOpen
  \bibfield  {author} {\bibinfo {author} {\bibfnamefont {M.}~\bibnamefont
  {{Tatsuuma}}}, \bibinfo {author} {\bibfnamefont {A.}~\bibnamefont
  {{Kataoka}}},\ and\ \bibinfo {author} {\bibfnamefont {H.}~\bibnamefont
  {{Tanaka}}},\ }\bibfield  {title} {\bibinfo {title} {{Tensile Strength of
  Porous Dust Aggregates}},\ }\href {https://doi.org/10.3847/1538-4357/ab09f7}
  {\bibfield  {journal} {\bibinfo  {journal} {\apj}\ }\textbf {\bibinfo
  {volume} {874}},\ \bibinfo {eid} {159} (\bibinfo {year} {2019})}\BibitemShut
  {NoStop}%
\bibitem [{\citenamefont {{Tatsuuma}}\ \emph {et~al.}(2023)\citenamefont
  {{Tatsuuma}}, \citenamefont {{Kataoka}}, \citenamefont {{Okuzumi}},\ and\
  \citenamefont {{Tanaka}}}]{2023ApJ...953....6T}%
  \BibitemOpen
  \bibfield  {author} {\bibinfo {author} {\bibfnamefont {M.}~\bibnamefont
  {{Tatsuuma}}}, \bibinfo {author} {\bibfnamefont {A.}~\bibnamefont
  {{Kataoka}}}, \bibinfo {author} {\bibfnamefont {S.}~\bibnamefont
  {{Okuzumi}}},\ and\ \bibinfo {author} {\bibfnamefont {H.}~\bibnamefont
  {{Tanaka}}},\ }\bibfield  {title} {\bibinfo {title} {{Formulating Compressive
  Strength of Dust Aggregates from Low to High Volume Filling Factors with
  Numerical Simulations}},\ }\href {https://doi.org/10.3847/1538-4357/acdf43}
  {\bibfield  {journal} {\bibinfo  {journal} {\apj}\ }\textbf {\bibinfo
  {volume} {953}},\ \bibinfo {eid} {6} (\bibinfo {year} {2023})}\BibitemShut
  {NoStop}%
\bibitem [{Note1()}]{Note1}%
  \BibitemOpen
  \bibinfo {note} {We note that an equivalent equation was derived in
  Ref.~\cite {CHEN2023118742} in a smart way.}\BibitemShut {Stop}%
\bibitem [{\citenamefont {{Tanaka}}\ \emph {et~al.}(2012)\citenamefont
  {{Tanaka}}, \citenamefont {{Wada}}, \citenamefont {{Suyama}},\ and\
  \citenamefont {{Okuzumi}}}]{2012PThPS.195..101T}%
  \BibitemOpen
  \bibfield  {author} {\bibinfo {author} {\bibfnamefont {H.}~\bibnamefont
  {{Tanaka}}}, \bibinfo {author} {\bibfnamefont {K.}~\bibnamefont {{Wada}}},
  \bibinfo {author} {\bibfnamefont {T.}~\bibnamefont {{Suyama}}},\ and\
  \bibinfo {author} {\bibfnamefont {S.}~\bibnamefont {{Okuzumi}}},\ }\bibfield
  {title} {\bibinfo {title} {{Growth of Cosmic Dust Aggregates and
  Reexamination of Particle Interaction Models}},\ }\href
  {https://doi.org/10.1143/PTPS.195.101} {\bibfield  {journal} {\bibinfo
  {journal} {Progress of Theoretical Physics Supplement}\ }\textbf {\bibinfo
  {volume} {195}},\ \bibinfo {pages} {101} (\bibinfo {year}
  {2012})}\BibitemShut {NoStop}%
\bibitem [{\citenamefont {{Krijt}}\ \emph {et~al.}(2013)\citenamefont
  {{Krijt}}, \citenamefont {{G{\"u}ttler}}, \citenamefont {{Hei{\ss}elmann}},
  \citenamefont {{Dominik}},\ and\ \citenamefont
  {{Tielens}}}]{2013JPhD...46Q5303K}%
  \BibitemOpen
  \bibfield  {author} {\bibinfo {author} {\bibfnamefont {S.}~\bibnamefont
  {{Krijt}}}, \bibinfo {author} {\bibfnamefont {C.}~\bibnamefont
  {{G{\"u}ttler}}}, \bibinfo {author} {\bibfnamefont {D.}~\bibnamefont
  {{Hei{\ss}elmann}}}, \bibinfo {author} {\bibfnamefont {C.}~\bibnamefont
  {{Dominik}}},\ and\ \bibinfo {author} {\bibfnamefont {A.~G.~G.~M.}\
  \bibnamefont {{Tielens}}},\ }\bibfield  {title} {\bibinfo {title} {{Energy
  dissipation in head-on collisions of spheres}},\ }\href
  {https://doi.org/10.1088/0022-3727/46/43/435303} {\bibfield  {journal}
  {\bibinfo  {journal} {Journal of Physics D: Applied Physics}\ }\textbf
  {\bibinfo {volume} {46}},\ \bibinfo {eid} {435303} (\bibinfo {year}
  {2013})}\BibitemShut {NoStop}%
\bibitem [{\citenamefont {{Arakawa}}\ and\ \citenamefont
  {{Krijt}}(2021)}]{2021ApJ...910..130A}%
  \BibitemOpen
  \bibfield  {author} {\bibinfo {author} {\bibfnamefont {S.}~\bibnamefont
  {{Arakawa}}}\ and\ \bibinfo {author} {\bibfnamefont {S.}~\bibnamefont
  {{Krijt}}},\ }\bibfield  {title} {\bibinfo {title} {{On the Stickiness of
  CO$_{2}$ and H$_{2}$O Ice Particles}},\ }\href
  {https://doi.org/10.3847/1538-4357/abe61d} {\bibfield  {journal} {\bibinfo
  {journal} {\apj}\ }\textbf {\bibinfo {volume} {910}},\ \bibinfo {eid} {130}
  (\bibinfo {year} {2021})}\BibitemShut {NoStop}%
\bibitem [{Note2()}]{Note2}%
  \BibitemOpen
  \bibinfo {note} {Here we analyze all particle pairs with $x > 0$ as a post
  process analysis. Strictly speaking, two particles in contact do not detach
  at $x = 0$ but the contact breaks at $x = - {(9 / 16)}^{1/3}$ in the JKR
  model. However, in our compression simulation, the contribution of contacts
  with $x < 0$ should be negligibly small.}\BibitemShut {Stop}%
\bibitem [{\citenamefont {{Arakawa}}\ \emph
  {et~al.}(2019{\natexlab{a}})\citenamefont {{Arakawa}}, \citenamefont
  {{Takemoto}},\ and\ \citenamefont {{Nakamoto}}}]{2019PTEP.2019i3E02A}%
  \BibitemOpen
  \bibfield  {author} {\bibinfo {author} {\bibfnamefont {S.}~\bibnamefont
  {{Arakawa}}}, \bibinfo {author} {\bibfnamefont {M.}~\bibnamefont
  {{Takemoto}}},\ and\ \bibinfo {author} {\bibfnamefont {T.}~\bibnamefont
  {{Nakamoto}}},\ }\bibfield  {title} {\bibinfo {title} {{Geometrical structure
  and thermal conductivity of dust aggregates formed via ballistic
  cluster-cluster aggregation}},\ }\href {https://doi.org/10.1093/ptep/ptz102}
  {\bibfield  {journal} {\bibinfo  {journal} {Progress of Theoretical and
  Experimental Physics}\ }\textbf {\bibinfo {volume} {2019}},\ \bibinfo {eid}
  {093E02} (\bibinfo {year} {2019}{\natexlab{a}})}\BibitemShut {NoStop}%
\bibitem [{\citenamefont {{Haile}}(1997)}]{haile1997molecular}%
  \BibitemOpen
  \bibfield  {author} {\bibinfo {author} {\bibfnamefont {J.}~\bibnamefont
  {{Haile}}},\ }\href {https://books.google.co.jp/books?id=ncPQwAEACAAJ} {\emph
  {\bibinfo {title} {{Molecular Dynamics Simulation: Elementary Methods}}}}\
  (\bibinfo  {publisher} {Wiley},\ \bibinfo {year} {1997})\BibitemShut
  {NoStop}%
\bibitem [{\citenamefont {{O'Sullivan}}(2011)}]{o2011particulate}%
  \BibitemOpen
  \bibfield  {author} {\bibinfo {author} {\bibfnamefont {C.}~\bibnamefont
  {{O'Sullivan}}},\ }\href@noop {} {\emph {\bibinfo {title} {{Particulate
  discrete element modelling: a geomechanics perspective}}}}\ (\bibinfo
  {publisher} {CRC Press},\ \bibinfo {year} {2011})\BibitemShut {NoStop}%
\bibitem [{\citenamefont {{Arakawa}}\ \emph
  {et~al.}(2019{\natexlab{b}})\citenamefont {{Arakawa}}, \citenamefont
  {{Tatsuuma}}, \citenamefont {{Sakatani}},\ and\ \citenamefont
  {{Nakamoto}}}]{2019Icar..324....8A}%
  \BibitemOpen
  \bibfield  {author} {\bibinfo {author} {\bibfnamefont {S.}~\bibnamefont
  {{Arakawa}}}, \bibinfo {author} {\bibfnamefont {M.}~\bibnamefont
  {{Tatsuuma}}}, \bibinfo {author} {\bibfnamefont {N.}~\bibnamefont
  {{Sakatani}}},\ and\ \bibinfo {author} {\bibfnamefont {T.}~\bibnamefont
  {{Nakamoto}}},\ }\bibfield  {title} {\bibinfo {title} {{Thermal conductivity
  and coordination number of compressed dust aggregates}},\ }\href
  {https://doi.org/10.1016/j.icarus.2019.01.022} {\bibfield  {journal}
  {\bibinfo  {journal} {Icarus}\ }\textbf {\bibinfo {volume} {324}},\ \bibinfo
  {pages} {8} (\bibinfo {year} {2019}{\natexlab{b}})}\BibitemShut {NoStop}%
\bibitem [{\citenamefont {{Seizinger}}\ and\ \citenamefont
  {{Kley}}(2013)}]{2013A&A...551A..65S}%
  \BibitemOpen
  \bibfield  {author} {\bibinfo {author} {\bibfnamefont {A.}~\bibnamefont
  {{Seizinger}}}\ and\ \bibinfo {author} {\bibfnamefont {W.}~\bibnamefont
  {{Kley}}},\ }\bibfield  {title} {\bibinfo {title} {{Bouncing behavior of
  microscopic dust aggregates}},\ }\href
  {https://doi.org/10.1051/0004-6361/201220946} {\bibfield  {journal} {\bibinfo
   {journal} {Astronomy \& Astrophysics}\ }\textbf {\bibinfo {volume} {551}},\
  \bibinfo {eid} {A65} (\bibinfo {year} {2013})}\BibitemShut {NoStop}%
\bibitem [{\citenamefont {{Dominik}}\ and\ \citenamefont
  {{Tielens}}(1995)}]{1995PMagA..72..783D}%
  \BibitemOpen
  \bibfield  {author} {\bibinfo {author} {\bibfnamefont {C.}~\bibnamefont
  {{Dominik}}}\ and\ \bibinfo {author} {\bibfnamefont {A.~G.~G.~M.}\
  \bibnamefont {{Tielens}}},\ }\bibfield  {title} {\bibinfo {title}
  {{Resistance to rolling in the adhesive contact of two elastic spheres}},\
  }\href {https://doi.org/10.1080/01418619508243800} {\bibfield  {journal}
  {\bibinfo  {journal} {Philosophical Magazine, Part A}\ }\textbf {\bibinfo
  {volume} {72}},\ \bibinfo {pages} {783} (\bibinfo {year} {1995})}\BibitemShut
  {NoStop}%
\bibitem [{\citenamefont {{Dominik}}\ and\ \citenamefont
  {{Tielens}}(1996)}]{1996PMagA..73.1279D}%
  \BibitemOpen
  \bibfield  {author} {\bibinfo {author} {\bibfnamefont {C.}~\bibnamefont
  {{Dominik}}}\ and\ \bibinfo {author} {\bibfnamefont {A.~G.~G.~M.}\
  \bibnamefont {{Tielens}}},\ }\bibfield  {title} {\bibinfo {title}
  {{Resistance to sliding on atomic scales in the adhesive contact of two
  elastic spheres}},\ }\href {https://doi.org/10.1080/01418619608245132}
  {\bibfield  {journal} {\bibinfo  {journal} {Philosophical Magazine, Part A}\
  }\textbf {\bibinfo {volume} {73}},\ \bibinfo {pages} {1279} (\bibinfo {year}
  {1996})}\BibitemShut {NoStop}%
\bibitem [{\citenamefont {{Chen}}\ \emph {et~al.}(2023)\citenamefont {{Chen}},
  \citenamefont {{Krengel}}, \citenamefont {{Nishiura}}, \citenamefont
  {{Furuichi}},\ and\ \citenamefont {{Matuttis}}}]{CHEN2023118742}%
  \BibitemOpen
  \bibfield  {author} {\bibinfo {author} {\bibfnamefont {J.}~\bibnamefont
  {{Chen}}}, \bibinfo {author} {\bibfnamefont {D.}~\bibnamefont {{Krengel}}},
  \bibinfo {author} {\bibfnamefont {D.}~\bibnamefont {{Nishiura}}}, \bibinfo
  {author} {\bibfnamefont {M.}~\bibnamefont {{Furuichi}}},\ and\ \bibinfo
  {author} {\bibfnamefont {H.-G.}\ \bibnamefont {{Matuttis}}},\ }\bibfield
  {title} {\bibinfo {title} {{A force^^e2^^80^^93displacement relation based on
  the JKR theory for DEM simulations of adhesive particles}},\ }\href
  {https://doi.org/https://doi.org/10.1016/j.powtec.2023.118742} {\bibfield
  {journal} {\bibinfo  {journal} {Powder Technology}\ }\textbf {\bibinfo
  {volume} {427}},\ \bibinfo {pages} {118742} (\bibinfo {year}
  {2023})}\BibitemShut {NoStop}%
\end{thebibliography}%

\end{document}